# Data-Driven Subsynchronous Oscillation Suppression for Renewable Energy Integrated Power Systems Based on Koopman Operator

Zihan Wang, *Member*, *IEEE*, Ziyang, Huang, *Member*, *IEEE*, Xiaonan Zhang, *Student Member*, *IEEE*, Gengyin Li, *Member*, *IEEE*, *Senior Member*, *CSEE*, and Le Zheng, *Member*, *IEEE*

***Abstract*—Recently, subsynchronous oscillations (SSOs) have emerged frequently worldwide, with the high penetration of renewable power generation in modern power systems. The SSO introduced by renewables has become a prominent new stability problem, seriously threatening the stable operation of systems. This paper proposes a data-driven dynamic optimal controller for renewable energy integrated power systems, to suppress SSOs with the control of renewables. The challenges of the controller design are the nonlinearity, complexity and hard accessibility of the system models. Using Koopman operator, the system dynamics are accurately extracted from data and utilized to the linear model predictive control (MPC). Firstly, the globally linear representation of the system dynamics is obtained by lifting, and the key states are selected as control signals by analyzing Koopman participation factors. Subsequently, augmented with control terms, the Koopman linear parameter-varying predictor of the nonlinear controlled system is constructed, acting as a more accurate predictor. Finally, using the computationally efficient linear MPC within the lifted Koopman space, the proposed controller computes control signals online in a moving horizon fashion. Case studies show that the proposed fully data-driven SSO suppression controller is effective, adaptive and robust in various conditions, surpassing other controllers with reliable control performance.**

## I. Introduction

THE modern power system has experienced a high penetration of renewable power generation such as wind power and solar power, due to the pressure of energy transition and net-zero carbon footprint. With the integration of bulk renewable power generators (RPGs) into the power system via flexibly controlled converters, subsynchronous oscillations (SSOs) caused by the interaction of converters and the AC grid are very prominent [1], [2]. The SSO introduced by RPGs has become an undesirable stability problem in modern power systems. For instance, in 2015, the wind power SSOs at 30 Hz occurred in Xinjiang China [1]. These oscillations propagated to the main grid and stimulated the protection relay of a 600-MW thermal power plant 48 km away, causing it to trip. In 2021, 22-Hz SSOs related to a solar photovoltaic (PV) farm were reported in eastern U.S., and instantaneous currents and voltages exhibited components at 38 Hz and 82 Hz [2]. The SSOs involve a large number of electrical components, create instabilities at frequencies below the nominal system frequency, and seriously threaten the stable operation of renewable energy integrated (REI) power systems. In real-world power systems, as the amplitude of the divergent SSO exceeds the small-signal range, the nonlinear characteristics of the system will emerge and dominate the oscillation dynamics. Most of the existing frequency-domain modeling and stability analysis methods are in the small-signal sense and thus unable to consider the impact of such nonlinearities [3]. What's more, the REI power systems are encountering increasing levels of nonlinearity, scale and dimensionality [4], [5], [6], posing more challenges and difficulties to the analysis and suppression of SSOs. Therefore, it is necessary to design an effective and advanced controller to suppress SSOs for REI power systems.

In general, there are two methods to design SSO suppression controllers: model-based method and data-driven method. The former is a conventional technique that often adopts the principle of phase compensation using the local linearization model of the system [7]. References [1] and [8] respectively attach supplementary damping controllers to the converter control circuit of a type-3 or type-4 wind turbine generator (WTG), which mitigate SSOs by compensating for mode phase and providing positive damping. In order to suppress SSOs under multi-operation conditions, an improved supplementary damping controller is proposed for a wind power generation system by adopting the particle swarm optimization algorithm in [9]. However, the design of this kind of controllers depends on the detailed mathematical model, which is hard to obtain for real-world power systems due to commercial privacy concerns, high nonlinearity, high complexity, or large scale. On the other hand, the operation condition variation may deteriorate the control performance since the parameters are tuned at the fixed operation point through the small-signal linearization.

Meanwhile, the data-driven method avoids the dependence of detailed mathematical models of power systems and has led to the development of numerous data-driven supplementary controllers. However, among these controllers, some rely heavily on the artificial intelligence algorithms, which lack interpretability and physical insight, and have kept the method from practical applications [4].

The Koopman operator (KO) theory integrates the flexibility and power of the data-driven method with more structured and insightful physical features derived from domain expertise. As a promising advance in data-driven analysis and control, the KO theory originated in the 1930s and has gained a growing

This work was supported by the National Key Research and Development Program of China (Response-Driven Intelligent Enhanced Analysis and Control for Bulk Power System Stability) under Grant 2021YFB2400800.

Z. Wang, Z. Huang, X. Zhang, G. Li and L. Zheng (corresponding author, e-mail: zhengl20@ncepu.edu.cn) are with State Key Laboratory of Alternate Electrical Power System with Renewable Energy Sources, North China Electric Power University, Beijing, 102206, China.

interest in the nonlinear community [10]-[18] over the last decade. This theory achieves the global linearization by lifting the nonlinear dynamical system with an infinite dimensional linear operator [10]. Mezić argues that the eigenfunctions of KO reveal important global geometric properties of the underlying nonlinear system [11]-[15]. For practical purposes, the finite dimensional approximations are obtained by numerical methods, like Dynamic Mode Decomposition (DMD) [16], Extended DMD (EDMD) [17], [18], etc. The KO theory shows the compatibility with computationally efficient linear control techniques such as linear quadratic regulator (LQR) and model predictive control (MPC) in [10]-[14]. The finite data-driven approximation of KO results in a class of linear predictors, useful for formulating linear MPC of nonlinear dynamical systems with reduced computational complexity. Moreover, the KO-based MPC has been applied in power systems [12]. Utilizing the KO-based MPC, a stabilization controller is proposed for enhancing transient stability [13], and a wind farm frequency controller is developed to provide frequency support [14].

While the application of KO theory in control is promising, there are still challenges in the implementation process. Among them, considering that the approximation error of the KO is inevitable, how to minimize the operator approximation error for controlled systems is the issue addressed in this paper. The control performance of data-driven controllers is sensitive to the identification accuracy, and so is the KO-based MPC. Despite the global linearity in the space of Koopman observables, linearity of observables does not imply linearity with respect to the control input [12]. A linear time invariant (LTI) Koopman form is generally assumed in EDMD, which facilitates the use of MPC [11]. However, this assumption is insufficient to capture the underlying dynamics of the nonlinear controlled systems. Few references have discussed whether the input matrix is invariant in the KO lifted form. Therefore, an accurate state predictor needs to be constructed, with which a well-designed MPC strategy enables the data-driven controller to suppress SSOs with more reliable control performance.

For the design of a data-driven SSO suppression controller in REI power systems, this paper addresses the following problems based on measurement data: 1) how to select the control signals of the controller, which can exploit the targeted control of key state variables in the dominant mode; 2) how to identify the nonlinear controlled system with inputs, that is, to construct an accurate predictor considering the effects of inputs on the system dynamics; 3) how to determine an effective control strategy based on the dynamic evolution form of the proposed state predictor.

In this paper, we propose a data-driven SSO suppression controller based on KO for REI power systems with the control of RPGs. Then, from measurement data, the selection of control signals, the identification of the controlled system and the determination of control strategy are implemented stage by stage. The major contributions are threefold:

1) A data-driven SSO suppression controller design framework is proposed. To the best of the authors' knowledge, this is the first comprehensive application of KO in SSO suppression controller design, fully harnessing the potential of KO in oscillation mode analysis and nonlinear dynamics characterization. Evolving around KO, three stages including signal selection, state prediction, and linear control utilization organically form the data-driven controller design framework.

2) A Koopman linear parameter-varying (KLPV) predictor in the globally linear representation is constructed, which enhances the prediction accuracy for SSO modes. The KLPV predictor considers the influence of inputs on the dynamics of the basis function during the Koopman linearization, accurately capturing the dynamic evolution of nonlinear controlled REI power systems.

3) A fully data-driven dynamic optimal control strategy is developed for REI power systems, which can suppress SSOs without the reliance on system topology and model parameters. Incorporated with the KLPV predictor, this control strategy utilizes the linear MPC algorithm for nonlinear dynamical systems with reduced computational complexity. The proposed controller maintains effective control performance and possesses strong adaptability and robustness.

The rest of the paper is organized as follows: Section II gives a brief introduction of KO and recent advances in global linearization and modal analysis using the operator. Section III derives the KLPV predictor and proposes the data-driven SSO suppression controller. Section IV describes an illustrative example of a REI power system with weak grid SSOs. Section V verifies the effectiveness and advantages of the proposed controller. Section VI draws the conclusion of this work.

## II. Preliminaries on Koopman Operator

In the context of power systems, the electromechanical and electromagnetical dynamics can be described by differential equations defined in the state space, according to the implicit function theorem [19]. Considering the nonlinearity of differential equations and the difficulty they pose for system analysis and control, the KO provides an alternative description of system dynamics in a linear observable space through a lifting process [11]. In this section, we briefly introduce the KO theory, Koopman mode decomposition and modal participation factors for nonlinear dynamical systems.

### A. Koopman Operator Theory

The continuous-time representation of dynamical systems often shows up in physical systems, and the function $\boldsymbol{f}$ to describe the temporal evolution of state variables, representing the differential equations in the state space, is of the form

$$\dot{\boldsymbol{x}} = \boldsymbol{f}(\boldsymbol{x}) \tag{1}$$

where $\boldsymbol{x} \in \boldsymbol{\mathcal{M}}$ is an $n$-dimension vector of state variables, including generator rotor angles, q-axis generator voltages, etc. $\boldsymbol{\mathcal{M}}$ is a differentiable manifold, often given by $\boldsymbol{\mathcal{M}} = \mathbb{R}^n$.

The KO $\mathcal{K}$ is a linear operator that acts in the observable space through the composition:

$$\mathcal{K}g = g \circ \boldsymbol{F}^t \equiv g(\boldsymbol{F}^t) \tag{2}$$

where $\boldsymbol{F}^t(\boldsymbol{x}_0) = \boldsymbol{x}_0 + \int_0^t \boldsymbol{f}(\boldsymbol{x}(\boldsymbol{\tau}))d\boldsymbol{\tau}$ is the flow map of the dynamics in (1), and the observable $g: \mathbb{R}^n \to \mathbb{C}$ is a scalar-valued function defined in the state space.

Discrete-time dynamical systems are more general and practical, because this representation is more consistent with experimental measurements collected from dynamical systems [10]. In discrete time, all the input data are sampling evenly from the flow map $\boldsymbol{F}^t$, dynamical systems are given by

$$\boldsymbol{x}_{k+1} = \boldsymbol{H}(\boldsymbol{x}_k) \tag{3}$$

where $\boldsymbol{H}$: $\mathbb{R}^n \to \mathbb{R}^n$ is the map, $\boldsymbol{x}_k = \boldsymbol{x}(k\Delta T)$, and $\Delta T$ means the sampling period.

The discrete-time KO $\mathcal{K}_\mathrm{d}$, as a linear map in the space of observables $g$, is defined in the following form

$$\mathcal{K}_\mathrm{d} g = g \circ \boldsymbol{H} \equiv g(\boldsymbol{H}) \tag{4}$$

### B. Koopman Mode Decomposition and Participation Factors

Since the KO is linear, it is natural to study the Koopman mode decomposition and modal participation factors by its spectral properties, i.e., eigenvalues and eigenfunctions. Considering discrete-time dynamical systems given by (3), the eigenvalues $\mu_i$ and eigenfunctions $\varphi_i$ have the form

$$\mathcal{K}_\mathrm{d}\varphi_i(\boldsymbol{x}_k) = \mu_i \varphi_i(\boldsymbol{x}_k), \quad i = 1, 2, \ldots \tag{5}$$

Given a vector-valued observable $\boldsymbol{g}$, if all the observables of $\boldsymbol{g}$ lie within the span of eigenfunctions $\varphi_i$, we have

$$\boldsymbol{g}(\boldsymbol{x}_k) = \sum_{i=1}^{\infty} \varphi_i(\boldsymbol{x}_k)\phi_i = \sum_{i=1}^{\infty} \varphi_i(\boldsymbol{x}_0)\phi_i \mu_i^k \tag{6}$$

where the vectors $\phi_i$ are defined as Koopman modes of the dynamical system, which can accurately describe the modal dynamics of dynamical systems [15], [18].

Since the KO is infinite dimensional, it is necessary to consider finite-dimensional approximations for practical applications. In order to approximate the finite-dimensional truncation of the KO directly from time series data, EDMD algorithm is adopted as proposed in [17]. EDMD requires two prerequisites. One is a dataset of snapshot pairs sampled from the state space, $\boldsymbol{X} = [\boldsymbol{x}_0, \boldsymbol{x}_1, \cdots, \boldsymbol{x}_{D-1}]$ and $\boldsymbol{Y} = [\boldsymbol{x}_1, \boldsymbol{x}_2, \cdots, \boldsymbol{x}_D]$, where $D$ is the total number of snapshots and $\boldsymbol{X}, \boldsymbol{Y} \in \mathbb{R}^{n\times D}$. The other is an $m$-dimension vector of scalar-valued function called the basis function, $\gamma(\boldsymbol{x})=[\gamma_1(\boldsymbol{x}), \gamma_2(\boldsymbol{x}), \cdots, \gamma_m(\boldsymbol{x})]^\mathrm{T}$, i.e., lifted states, where T denotes the transpose operator.

By using the basis function $\gamma(\boldsymbol{x})$, the system is lifted from the $n$-dimension state space to the $m$-dimension observable space:

$$\begin{aligned} \boldsymbol{X}_\mathrm{lift} &= [\gamma(\boldsymbol{x}_0), \gamma(\boldsymbol{x}_1), \cdots, \gamma(\boldsymbol{x}_{D-1})], \\ \boldsymbol{Y}_\mathrm{lift} &= [\gamma(\boldsymbol{x}_1), \gamma(\boldsymbol{x}_2), \cdots, \gamma(\boldsymbol{x}_D)] \end{aligned} \tag{7}$$

Then, a finite-dimensional approximation to the KO $\mathcal{K}_\mathrm{d}$ in a least square sense is introduced as

$$\boldsymbol{K}_\mathrm{d} = \boldsymbol{Y}_\mathrm{lift}\boldsymbol{X}_\mathrm{lift}^\dagger \tag{8}$$

where $\dagger$ means the Moore-Penrose pseudoinverse. The computational complexity of the least square problem is $\mathcal{O}(m^2 D)$, which depends on the dimension $m$ of the basis functions and the number $D$ of data snapshot pairs.

The associated eigenfunctions of the approximation $\boldsymbol{K}_\mathrm{d}$ are

$$\boldsymbol{\varphi}(\boldsymbol{x}_k) = \Xi\gamma(\boldsymbol{x}_k) \tag{9}$$

with $\Xi := [\xi_1, \xi_2, \cdots, \xi_m]^\mathrm{T}$, and $\xi_i$ is the $i$-th left eigenvector.

Given (9) and properties of the KO as well as its eigenvectors, the $i$-th eigenfunction has the evolution form

$$\varphi_i(\boldsymbol{x}_k) = \mu_i^k \varphi_i(\boldsymbol{x}_0) = \mu_i^k \xi_i \gamma(\boldsymbol{x}_0) = \mu_i^k \sum_{j=1}^{m} \xi_{ij}\gamma(\boldsymbol{x}_0) \tag{10}$$

The proof is shown in Appendix A, and this equation reveals the contributions of each component $\gamma_j(\boldsymbol{x}_0)$ of the initial lifted states $\gamma(\boldsymbol{x}_0)$ to the evolution of the $i$-th mode. By averaging the relative contribution of $\gamma(\boldsymbol{x}_0)$ in modes and evaluating the result at initial time [20], the state-in-mode Koopman participation factors (KPFs) for the $j$-th lifted state in the $i$-th mode of nonlinear systems are defined as

$$p_{ij} = \frac{\left(\mathrm{Re}\{\xi_{ij}\}\right)^2}{\left(\mathrm{Re}\{\xi_j\}\right)^\mathrm{T} \mathrm{Re}\{\xi_j\}} \tag{11}$$

Due to space limitation, more details about the data-driven KPFs can be found in [18]. The Koopman modes can provide a description of the modal characteristics of nonlinear models and KPFs can quantify the relative contribution of different state variables in the dominant mode by using measurement data. Consequently, the data-driven KPF analysis lays an important foundation for the selection of control signals in a controlled power system susceptible to oscillations.

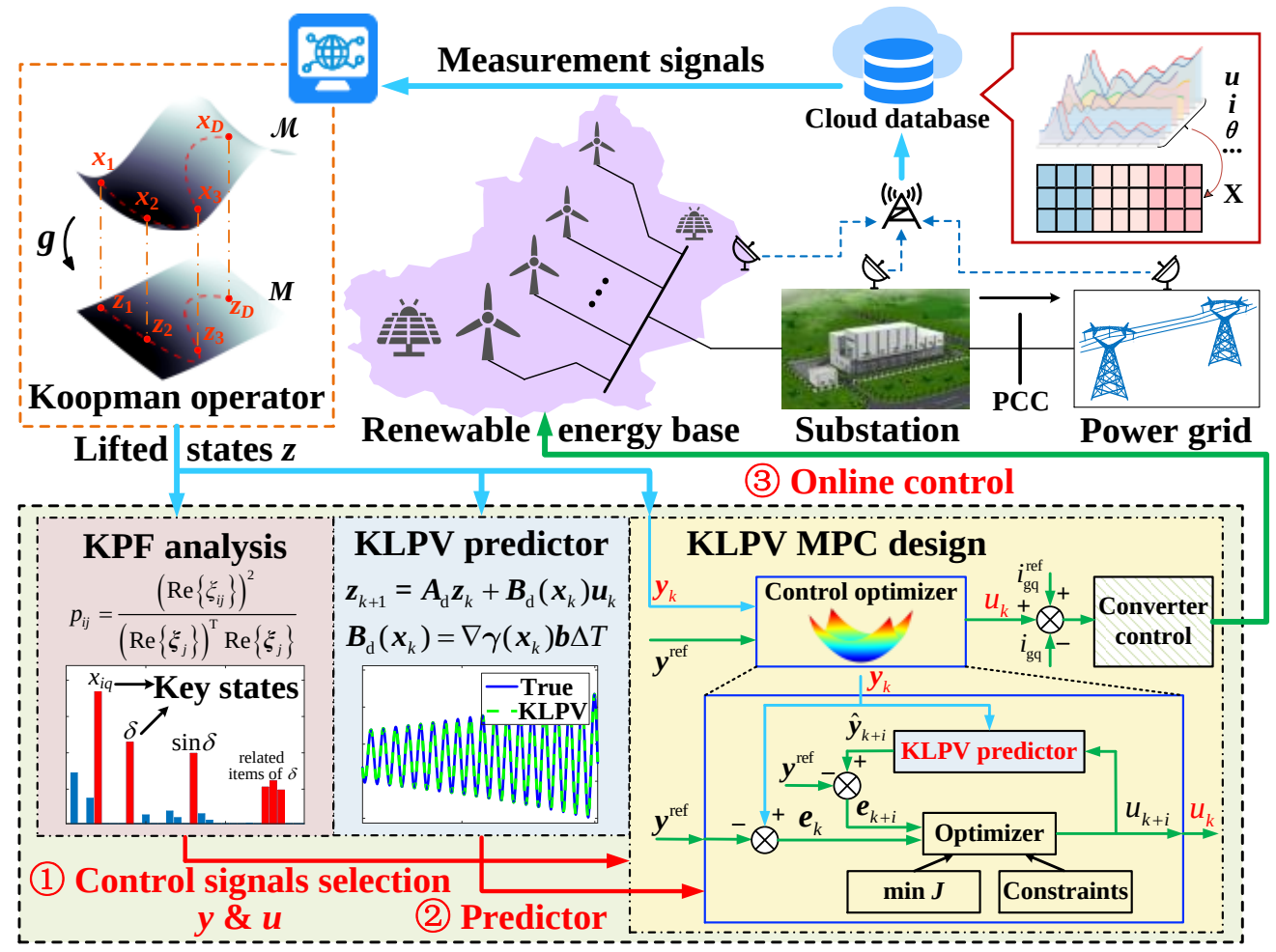

Fig. 1. Schematic of DSSOSC design framework for a controlled system. Blue arrows: transmission of measurement data; red arrows: transmission of key information in controller design; green arrows: transmission of control signals.

## III. Data-Driven SSO Suppression Controller Design

In this section, a data-driven subsynchronous oscillation suppression controller (DSSOSC) is designed for REI power systems, by using observed data from the system dynamics. The aim is to find an effective control algorithm to suppress persistent SSOs and to enhance the stability of power systems with the regulation of RPGs. In conjunction with KO, the KLPV predictor for the controlled system with inputs is derived, and the MPC in the KLPV form is utilized as the control strategy for the proposed data-driven controller.

### A. DSSOSC Design Framework for Controlled Systems

Fig. 1 shows the design framework of DSSOSC in REI power systems. The proposed controller DSSOSC serves as a supplementary module that is attached to the control cabinets of converters within RPGs. With the progressive development of communication, computation and control, once wide-area measurement signals are obtained, the DSSOSC can compute key mode information and control sequences. This enables the controller to regulate control references of RPGs online, and effectively suppress SSOs in the REI power system.

The controlled system is a REI power system susceptible to SSOs, and a simplified but representative system model is shown in the upper part of the schematic. According to the survey of real-world SSO events associated with WTGs and PVs [2], a weak grid interconnection with a high penetration of renewable power generation is viewed as a significant challenge to the system. For example, in a renewable energy base located in Xinjiang China, the renewable power generation has to be transmitted to the main power grid with a transformer substation through long-distance transmission lines. Therefore, the short circuit ratio of the REI power system is relatively low [1], which is in line with the challenge of renewable integration into the weak grid.

The remaining part of the schematic focuses on the selection of control signals, identification of the controlled system with inputs and determination of control strategy. Correspondingly, based on KO, the analysis of participation factors, the construction of state predictors, and the utilization of MPC formulation organically form the three main stages of the controller design. Elaborated in **Algorithm 1**, in the offline procedures, the DSSOSC design method utilizes massive data from cloud databases to implement the design of the first two stages, i.e., analyzing the KPFs of potential oscillation modes, and yielding a global KLPV predictor. In the online procedures, the method implements the third stage to iteratively compute control signals in a moving horizon fashion.

**Algorithm 1** DSSOSC design method

**Offline procedures:**

**1** Select basis functions $\gamma(\boldsymbol{x})$ and prepare datasets.

**2** Compute a finite-dimensional approximation $\boldsymbol{K}_\mathrm{d}$ in the globally linear representation of uncontrolled system.

**3** Compute state-in-mode KPFs of potential oscillation modes, sort descending the KPFs, and then select input and output signals of the controller, $\boldsymbol{y}_k$ and $\boldsymbol{u}_k$.

**4** Compute the state-varying input matrix $\boldsymbol{B}_\mathrm{d}(\boldsymbol{x}_k)$ with $\boldsymbol{B}_\mathrm{d}(\boldsymbol{x}_k)=\nabla\gamma(\boldsymbol{x}_k)\boldsymbol{b}\Delta T$ for controlled system with inputs $\boldsymbol{u}_k$.

**5** Construct the KLPV predictor of the controlled system by the extension of the control term $\boldsymbol{B}_\mathrm{d}(\boldsymbol{x}_k)\boldsymbol{u}_k$.

**Online procedures:**

At each discrete-time step $k = 1, 2, \cdots$

**1** Measure $\boldsymbol{x}_k$ and lifted states $\boldsymbol{z}_k$, and initialize $\boldsymbol{B}_\mathrm{d}(\boldsymbol{x}_k)$.

**2** Solve the MPC optimizer in the convex quadratic programming form, and obtain the optimal control sequences $(\boldsymbol{u}_i^*)_{i=0}^{N_\mathrm{p}-1}$.

**3** Apply the first element $\boldsymbol{u}_0^*$ to the controlled system in the form of the KLPV predictor.

## B. Koopman Linear Parameter-Varying Predictors

Linearity of Koopman observables does not imply the linearity with respect to the control input in the controlled systems [12]. When control signals have been selected and added to the controlled systems, the construction of Koopman form requires introducing the influence of inputs on the basis function. This influence is characterized as a state-dependent input matrix in the Koopman linear representation, hence the Koopman form can be interpreted as a linear parameter-varying model. Below the KLPV predictors are derived for controlled systems with inputs from measurement data.

### 1) Uncontrolled Systems in the Observable Space

Firstly, an uncontrolled dynamical system is lifted in the Koopman observable space, giving time-domain trajectories of uncontrolled systems in the continuous and discrete representations.

The continuous-time lifted dynamics of the unactuated, autonomous system of the form (1) can be represented as

$$\dot{\gamma}(\boldsymbol{x}) = \boldsymbol{A}\gamma(\boldsymbol{x}) \tag{12}$$

where $\boldsymbol{A}$ is a state-transition matrix, approximate to the infinitesimal generator of one-parameter family of KOs [12].

The discrete-time lifted uncontrolled dynamics, which has been explained in EDMD, can be written as

$$\gamma(\boldsymbol{x}_{k+1}) = \boldsymbol{A}_\mathrm{d}\gamma(\boldsymbol{x}_k) \tag{13}$$

where the state-transition matrix $\boldsymbol{A}_\mathrm{d}$ is an finite-dimensional approximation of the KO, equal to $\boldsymbol{K}_\mathrm{d}$, which can be computed with a least square sense (8) from data. The $\boldsymbol{A}$ and $\boldsymbol{A}_\mathrm{d}$ satisfies the relation [12]:

$$\boldsymbol{A}\gamma(\boldsymbol{x}_0) = \lim_{t\to 0}\frac{\boldsymbol{A}_\mathrm{d}\gamma(\boldsymbol{x}_0) - \gamma(\boldsymbol{x}_0)}{t} \tag{14}$$

### 2) Controlled Systems with Inputs in the Observable Space

Next, how the inputs added to controlled systems affects the dynamics of the basis function is analyzed. Furthermore, KLPV predictors are constructed that are suitable for control purposes in the observable space.

Consider a control-affine system by decoupling contributions of uncontrolled and input-actuated dynamics:

$$\dot{\boldsymbol{x}} = \boldsymbol{f}_\mathrm{c}(\boldsymbol{x}, \boldsymbol{0}) + \boldsymbol{h}_\mathrm{c}(\boldsymbol{x}, \boldsymbol{u}) = \boldsymbol{f}(\boldsymbol{x}) + \boldsymbol{b}(\boldsymbol{x})\boldsymbol{u} \tag{15}$$

where $\boldsymbol{b}(\boldsymbol{x}) \in \mathbb{R}^{n\times q}$ is the input matrix function, and $\boldsymbol{u} \in \mathbb{R}^q$ is the input vector of the controlled system. The functions $\boldsymbol{f}_\mathrm{c}$ and $\boldsymbol{h}_\mathrm{c}$ represent contributions of the autonomous and input related dynamics, respectively.

By applying the chain rule, based on the uncontrolled dynamics (1) and (12), the continuous-time extension of the Koopman form with inputs can be augmented as

$$\dot{\gamma}(\boldsymbol{x}) = \nabla\gamma(\boldsymbol{x})\big(\boldsymbol{f}(\boldsymbol{x}) + \boldsymbol{b}(\boldsymbol{x})\boldsymbol{u}\big) = \boldsymbol{A}\gamma(\boldsymbol{x}) + \boldsymbol{B}(\boldsymbol{x})\boldsymbol{u} \tag{16}$$

with the input matrix function $\boldsymbol{B}(\boldsymbol{x}) := \nabla\gamma(\boldsymbol{x})\boldsymbol{b}(\boldsymbol{x})$ .

In discrete time, the chain rule can no longer be applied. After discretizing the controlled dynamics using the Euler method and with the relation between $\boldsymbol{A}$ and $\boldsymbol{A}_\mathrm{d}$ in (14), the extension of the Koopman form can be written as

$$\begin{aligned}\gamma(\boldsymbol{x}_{k+1}) &= \gamma(\boldsymbol{x}_k) + \int_{k\Delta T}^{(k+1)\Delta T}\big(\boldsymbol{A}\gamma(\boldsymbol{x}(\tau)) + \boldsymbol{B}(\boldsymbol{x}(\tau))\boldsymbol{u}(\tau)\big)d\tau \\ &= \gamma(\boldsymbol{x}_k) + \int_{k\Delta T}^{(k+1)\Delta T}\boldsymbol{A}\gamma(\boldsymbol{x}(\tau))d\tau + \int_{k\Delta T}^{(k+1)\Delta T}\boldsymbol{B}(\boldsymbol{x}(\tau))\boldsymbol{u}(\tau)d\tau \\ &= \boldsymbol{A}_\mathrm{d}\gamma(\boldsymbol{x}_k) + \boldsymbol{B}_\mathrm{d}(\boldsymbol{x}_k)\boldsymbol{u}_k\end{aligned} \tag{17}$$

with the input matrix function $\boldsymbol{B}_\mathrm{d}(\boldsymbol{x}_k) := \nabla\gamma(\boldsymbol{x}_k)\boldsymbol{b}(\boldsymbol{x}_k)\Delta T$ .

Therefore, KLPV predictors in the observable space have been obtained as (16) and (17). Note that the input matrix in KLPV predictors is state-varying, different from the constant input matrix in LTI predictors [11] that may neglect the influence of inputs on the KO basis function.

***Remark 1***: The KLPV predictors are suitable for control affine design methodologies such as MPC [21], and the design of the proposed controller adopts the same control law. In order to facilitate the implementation of the physical structure of the supplementary controller, linear inputs are added to the controlled system with a particular case $\boldsymbol{b}(\boldsymbol{x})=\boldsymbol{b}$.

***Remark 2***: The designed supplementary controller has clear actuated positions of the selected control signals, which directly determines whether each input affects related states or not, namely the correlation between inputs and states. Thus, $\boldsymbol{b}$ is usually considered known in the artificial design of the supplementary controller, and the input matrix in KLPV predictors satisfies $\boldsymbol{B}_{\mathrm{d}}(\boldsymbol{x}_k)=\nabla\gamma(\boldsymbol{x}_k)\boldsymbol{b}\Delta T$. If $\boldsymbol{b}$ is unknown due to uncertainty of parameters and difficulty in determining actuated positions, the KLPV predictor can be adjusted by a lifted bilinear form as $\gamma(\boldsymbol{x}_{k+1})=\boldsymbol{A}_{\mathrm{d}}\gamma(\boldsymbol{x}_k)+\sum_{i=1}^{q}\boldsymbol{\beta}_i\gamma(\boldsymbol{x}_k)\boldsymbol{u}_i(\boldsymbol{x}_k)$, where $\boldsymbol{u}_i$ is the $i$-th component of $\boldsymbol{u}$. As discussed in [12],[22] , assuming $\nabla\gamma\cdot\boldsymbol{b}\in span\{\gamma\}$ and $\nabla\gamma\cdot\boldsymbol{b}_i=\boldsymbol{\beta}_i\cdot\gamma$ with $\boldsymbol{b}_i$ being the $i$-th column of $\boldsymbol{b}$, the KLPV predictors of the bilinear form can be obtained. Then, the input matrix satisfies $\boldsymbol{B}_{\mathrm{d}}(\boldsymbol{x}_k)=[\boldsymbol{\beta}_1\gamma(\boldsymbol{x}_k), \boldsymbol{\beta}_2\gamma(\boldsymbol{x}_k), \cdots, \boldsymbol{\beta}_q\gamma(\boldsymbol{x}_k)]$, and from datasets of controlled systems, coefficient matrices $\boldsymbol{\beta}_i\in\mathbb{R}^{m\times m}$ can be computed by solving the least square optimization problem with $\boldsymbol{z}_k=\gamma(\boldsymbol{x}_k)$ as

$$\min_{\boldsymbol{A}_{\mathrm{d}},\boldsymbol{\beta}_1,\ldots,\boldsymbol{\beta}_q}\sum_{k=0}^{D-1}\left\|\boldsymbol{z}_{k+1}-(\boldsymbol{A}_{\mathrm{d}}\boldsymbol{z}_k+\sum_{i=1}^{q}\boldsymbol{\beta}_i\boldsymbol{z}_k\boldsymbol{u}_k(i))\right\|^2 \tag{18}$$

Summarizing, with augmenting the uncontrolled system with inputs, the KLPV predictor is constructed in (17) from a set of experimental data, where $\boldsymbol{A}_{\mathrm{d}}$ is computed with a least square sense (8) and $\boldsymbol{B}_{\mathrm{d}}(\boldsymbol{x}_k)$ is derived as explained in ***Remark 2***. This predictor can predict the time-domain trajectories of controlled systems, and provide global accuracy of nonlinear dynamic evolution for the data-driven controller design.

### *C. Model Predictive Control in the KLPV Form*

With the selected control signals and the KLPV predictor identified from measurements, the control strategy of DSSOSC proposed here utilizes the MPC formulation. MPC, which roots in optimal control, optimally chooses a sequence of control inputs online in a moving horizon fashion when forecasting dynamical system behavior [21].

To suppress SSOs, the KLPV MPC computes a control sequence at time step $t_k$ by optimizing a quadratic cost function over the prediction horizon. Let $N_{\mathrm{p}}$ be the length of the prediction horizon, and the sequence of input and output values over the receding horizon from $t_k$ are denoted by $(\boldsymbol{u}_{i|k})_{i=0}^{N_{\mathrm{p}}-1}$ and $(\boldsymbol{y}_{i|k})_{i=0}^{N_{\mathrm{p}}}$ . For brevity, the subscript related to $k$ is omitted later in this subsection. At each time step, MPC designed for DSSOSC strategy to regulate the reference values of key states in the dominant mode of oscillations can be expressed as

$$\min_{(\boldsymbol{u}_i)_{i=0}^{N_{\mathrm{p}}-1}} J=\sum_{i=0}^{N_{\mathrm{p}}-1}\left(\left\|\boldsymbol{y}_i-\boldsymbol{y}^{\mathrm{ref}}\right\|_{\boldsymbol{Q}}^2+\left\|\boldsymbol{u}_i\right\|_{\boldsymbol{R}}^2\right)+\left\|\boldsymbol{z}_{N_{\mathrm{p}}}-\boldsymbol{z}^{\mathrm{ref}}\right\|_{\boldsymbol{P}}^2 \tag{19}$$

$$s.t.\quad \boldsymbol{z}_{i+1}=\boldsymbol{A}_{\mathrm{d}}\boldsymbol{z}_i+\boldsymbol{B}_{\mathrm{d}}(\boldsymbol{x}_i)\boldsymbol{u}_i,\qquad i=0,\ldots,N_{\mathrm{p}}-1 \tag{20}$$

$$\boldsymbol{B}_{\mathrm{d}}(\boldsymbol{x}_i)=\nabla\gamma(\boldsymbol{x}_i)\boldsymbol{b}\Delta T,\qquad i=0,\ldots,N_{\mathrm{p}}-1 \tag{21}$$

$$\boldsymbol{y}_i=\boldsymbol{C}\boldsymbol{z}_i,\qquad i=0,\ldots,N_{\mathrm{p}} \tag{22}$$

$$\boldsymbol{\beta}_{i,\min}\le\boldsymbol{E}_i^y\boldsymbol{y}_i+\boldsymbol{E}_i^u\boldsymbol{u}_i\le\boldsymbol{\beta}_{i,\max},\quad i=0,\ldots,N_{\mathrm{p}}-1 \tag{23}$$

$$\boldsymbol{\beta}_{N_{\mathrm{p}},\min}\le\boldsymbol{E}_{N_{\mathrm{p}}}^y\boldsymbol{y}_{N_{\mathrm{p}}}\le\boldsymbol{\beta}_{N_{\mathrm{p}},\max}, \tag{24}$$

$$\boldsymbol{z}_0=\gamma(\boldsymbol{x}_0). \tag{25}$$

where $J$ is a quadratic cost function, and the objective is to minimize the weighed values of oscillation amplitudes of system outputs and control inputs under SSOs. For a vector $\boldsymbol{y}$, we use $\|\boldsymbol{y}\|_Q^2$ to stand for $\boldsymbol{y}^{\mathrm{T}}\boldsymbol{Q}\boldsymbol{y}$, $\|\boldsymbol{y}\|$ to denote its Euclidean norm. Positive-semidefinite matrices $\boldsymbol{Q}$ and $\boldsymbol{R}$ denote state weight and control weight and the symmetric positive-definite matrix $\boldsymbol{P}$ means the weight of terminal cost. $\boldsymbol{y}^{\mathrm{ref}}$ and $\boldsymbol{z}^{\mathrm{ref}}$ denote the reference values of system outputs and lifted state variables. The matrix $\boldsymbol{C}$ characterizes the map from lifted state variables $\boldsymbol{z}_i$ to system outputs $\boldsymbol{y}_i$. The matrices $\boldsymbol{E}_i^y$ and $\boldsymbol{E}_i^u$ and the vector $\boldsymbol{\beta}_{i,\max}$ define polyhedral constraints on the input and output values of the system. Note that outputs $\boldsymbol{y}_i$ of the controlled system, also serving as input signals of the DSSOSC, are the key states in the dominant mode selected through the KPF analysis. Inputs $\boldsymbol{u}_i$ of the controlled system are optimized iteratively and attached to positions of the output reference $\boldsymbol{y}^{\mathrm{ref}}$.

The KLPV predictor makes the MPC optimization problem non-convex due to the state-dependent input matrix $\boldsymbol{B}_{\mathrm{d}}(\boldsymbol{x}_k)$. Therefore, in order to effectively introduce KLPV predictor into MPC constraints, we improve the current Koopman LTI (KLTI) MPC formulation [11] in the KLPV form. Namely, the KLPV predictor is used to capture the controlled behavior at each time step in the actual temporal evolution, while a linear approximation is proposed by fixing the initial input matrix over each prediction horizon.

The dynamics constraints of the KLPV form in (20) and (21) are then transformed into a linear form,

$$\boldsymbol{z}_{i+1}=\boldsymbol{A}_{\mathrm{d}}\boldsymbol{z}_i+\boldsymbol{B}_{\mathrm{d},0}\boldsymbol{u}_i, \tag{26}$$

where initial input matrix $\boldsymbol{B}_{\mathrm{d},0}$ obeys $\boldsymbol{B}_{\mathrm{d},0}=\nabla\gamma(\boldsymbol{x}_0)\boldsymbol{b}\Delta T$. The prediction error is introduced by the approximation of input matrix initialization over the prediction horizon, and as a convex quadratic program solving quickly enough, KLPV MPC can make up for it by sacrificing a certain amount of regulation cycle with iteration. Besides, the recursive feasibility, asymptotic stability and inherent robustness of this linear MPC based on KO can be demonstrated with [23] and [24], which lay the theoretical foundation for DSSOSC to stabilize SSOs in real data-driven applications.

Thus, based on KLPV MPC, the fully data-driven dynamic optimal controller DSSOSC has been designed to suppress SSOs for REI power systems with the control of RPGs. This controller is designed for nonlinear dynamical systems with computational complexity comparable to MPC controllers for linear dynamical systems with the same number of control inputs and states. The key advantage of KLPV MPC over KLTI MPC is its enhanced accuracy in identifying nonlinear dynamics, and KLPV MPC with improved identification accuracy can ensure better control performance accordingly. This is achieved in its ability to capture the influence of changing state variables on the input matrix at the beginning of each prediction horizon.

## IV. ILLUSTRATIVE EXAMPLE WITH WEAK GRID SSOS

For power systems integrated with bulk RPGs, the weak grid operation has led to various SSO events [2]. To capture the dynamic behavior of REI power systems under a range of conditions, an illustrative example of a REI power system susceptible to SSOs is derived from the literature [1], [25].

A renewable energy base is supposed to aggregate $n$ identical RPGs with grid-following converters, which are of the same capacity, structure, parameters, and connected to the same bus. Considering the key of SSOs lies in the grid-side converter [2], and assuming RPGs are in the same status, the grid-following converter as a controlled current source is used to analyze the oscillation characteristics of each RPG in the power system [1].

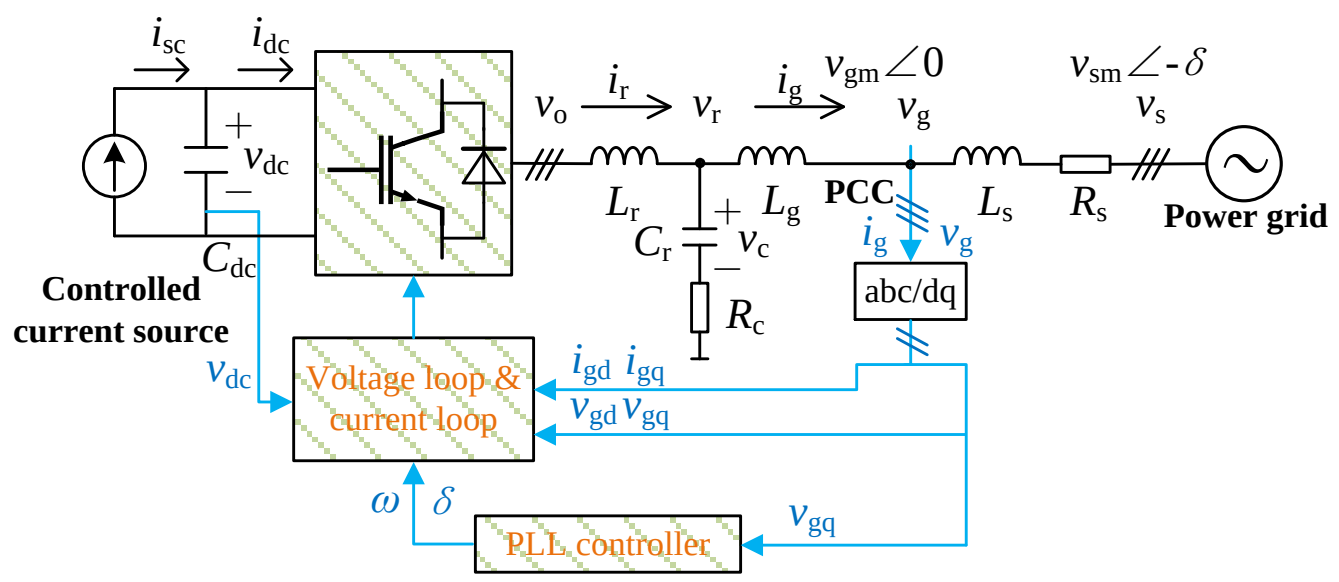

Fig. 2. Diagram of a REI power system susceptible to weak grid SSOs. Here, green shadows indicate that parameters of the gray-box modules cannot be obtained, and the network parameters of the system are also unknown.

A general schematic diagram of a REI power system with an LCL-type grid-following converter connected to a weak AC grid is shown in Fig. 2.

The REI power system is composed of a DC link, voltage loop, current loop, phase-locked loop (PLL), LCL filter and system network. The dynamics of the REI power system are detailed in Appendix B. When identifying this gray-box system based on KO, although its structure or parameters may not be available, we assume full access to the state vector $\boldsymbol{x}$ as

$$\boldsymbol{x}=[v_{dc},x_v,x_{id},x_{iq},v_{od},v_{oq},x_{\omega},\delta,i_{rd},i_{rq},v_{cd},v_{cq},i_{gd},i_{gq}]^{T} \quad (27)$$

where $v_{dc}$ is the DC link voltage, and $x_v$ is the state variable of the voltage loop. $x_{id}$ and $x_{iq}$ are the state variables of the current loop. $v_{od}$ and $v_{oq}$ are the output voltage variables. $x_{\omega}$ is the state variable of the PLL controller, and $\delta$ is the phase difference between the voltage of point of common coupling (PCC) $v_g$ and the voltage of power grid $v_s$, of which magnitudes are $v_{gm}$ and $v_{sm}$. Besides, $i_{rd}$ and $i_{rq}$ are the currents of converter-side inductance $L_r$. $v_{cd}$ and $v_{cq}$ are the voltages of capacitance $C_r$. $i_{gd}$ and $i_{gq}$ are the currents of grid-side inductance $L_g$.

These 14 state variables are expressed in the dq-frame, and can capture the physical features of the REI power system with weak grid SSOs. Some state variables not directly measured in practice can be estimated via a dynamic state estimator [26] (as a pre-filtering step). In this illustrative example, it is assumed that all state variables can be obtained, while the utilization of partial measurement data in larger real-world power systems will also be demonstrated in the next section of case studies.

The gray-box model of the REI power system conduces to provide the available knowledge from physical principles for predicting or characterizing the state variables, regardless of whether the system parameters are known or unknown. It also offers a physical prior knowledge for the KO basis function construction [4]. Therefore, the dynamic behavior of a REI power system with weak grid SSOs has been described for which we design controllers in the next section.

## V. CASE STUDIES

In this section, the effectiveness, adaptability and robustness of the proposed DSSOSC are evaluated by case studies. This controller is assessed through comparison in Matlab/Simulink on a computer with an Intel Core i7-11700F CPU at 2.50 GHz and a 64 GB RAM. The convex quadratic program in the linear MPC is solved by qpOASES [27] in Matlab 2023a.

### A. *Test Systems and Data-Driven Prerequisites*

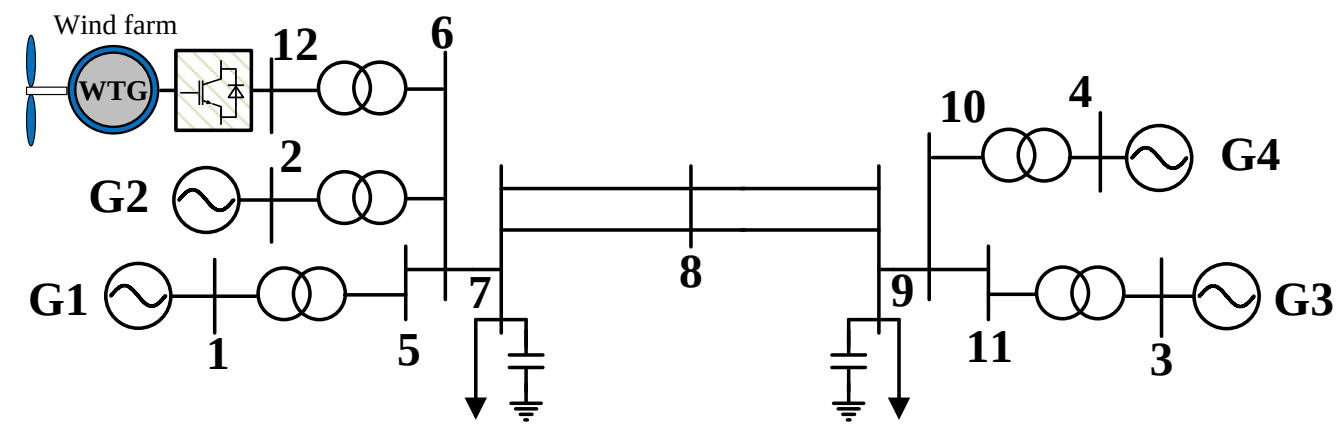

Fig. 3. Diagram of the modified REI Kundur two-area system.

The first test system is a typical REI power system with weak grid interconnection, as illustrated in Fig. 2. This test system is susceptible to SSOs, as a wind farm aggregating 700 type-4 WTGs generates electricity into the main power grid via 35 kV/110 kV and 110 kV/220 kV transformer and a long-distance line. The rated apparent power of a single WTG is 1.5 MVA, and all WTGs are collected into the collection bus boosted by a 0.62 kV/35 kV transformer. Each WTG is configured with an LCL-type grid-following converter as its grid-side converter. Fig. 3 exhibits the second test system, where a wind farm of 1000MW installed capacity is connected to Bus-6 via a long-distance transmission line in the REI Kundur two-area power system. The WTGs in the wind farm are also type-4, and parameters of four synchronous generators (SGs) are from [19].

Here is a typical operating condition suffering SSOs, where WTGs operate at low output levels, and at 0.5s, parameters of the grid structure suddenly change, equivalent to a step change of line reactance, which makes the AC grid even weaker. Changes in parameters of the grid structure can be caused by variations in parameters of the transformer substation, adjustments in the distance from WTGs to the collection bus, and changes in the operating status of the main power gird.

To achieve fast and accurate measurement of SSOs, considering the frequency range of SSOs, the reporting rate of measurement devices is set at 500 Hz. The data used to identify the KO of the REI power system consists of two datasets collected from the original model described in Section IV. One is uncontrolled system dataset with the control input $\boldsymbol{u}$=**0**, where 300 trajectories over 1000 snapshots were collected with randomized initial conditions. The other is the controlled system dataset, where 300 trajectories over 1000 snapshots were collected with a range of randomized initial conditions and inputs. For the choice of basis functions, we have made progress in accounting for the nonlinear function characteristics of the power system models in the KO basis function construction [4]. Based on this research, polynomial terms and trigonometric functions of the state variables are employed in the polynomial basis function up to third order, which contains terms with trigonometric functions such as sin$\delta$, cos$^2\delta$ and $i_{gq}$·sin$\delta$cos$\delta$, and treats sin$\delta$ and cos$\delta$ as first-order monomials. In addition, in the case study of the REI Kundur two-area system with SGs, rotor angle $\delta_{Gj}$, rotor speed deviation $\omega_{Gj}$, field voltage $E_{fdj}$, and real power injection $P_{Gj}$ ($j$=1,2,3,4) are added to the state matrix when collecting data.

### B. *Typical REI Power System with Weak Grid Interconnection*

#### 1) *Control Signal Selection*

When using a supplementary controller to suppress SSOs, the different input signals and installation locations (output signals) of the controller will affect its control performance, and thus need to be selected reasonably. The analysis of KPFs of nonlinear systems from measurements can quantify the impact of different state variables on the modes of SSOs.

TABLE I
DOMINANT MODE OF OSCILLATION FROM KOOPMAN MODE DECOMPOSITION IN THE FIRST TEST SYSTEM

| Mode Type | Eigenvalue | Frequency(Hz) | Damping Ratio(%) |
|---|---|---|---|
| SSO | 0.65±67.38i | 10.74 | -0.96 |

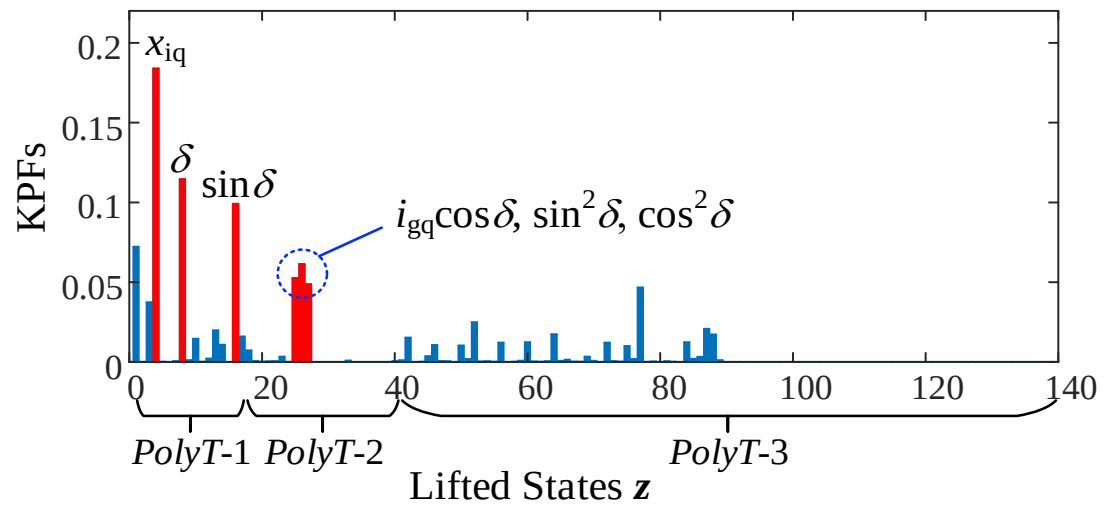

Fig. 4. State-in-mode Koopman participation factors of the first test system.

TABLE II
COMPONENT PARTICIPATION OF THE REI SYSTEM UNDER SSO MODE

| Components | Participation |
|---|---|
| DC Link | 0.0719 |
| Voltage Loop | 0.0604 |
| Current Loop | 0.3078 |
| PLL Controller | 0.3795 |
| LCL Filter and the Network | 0.1804 |

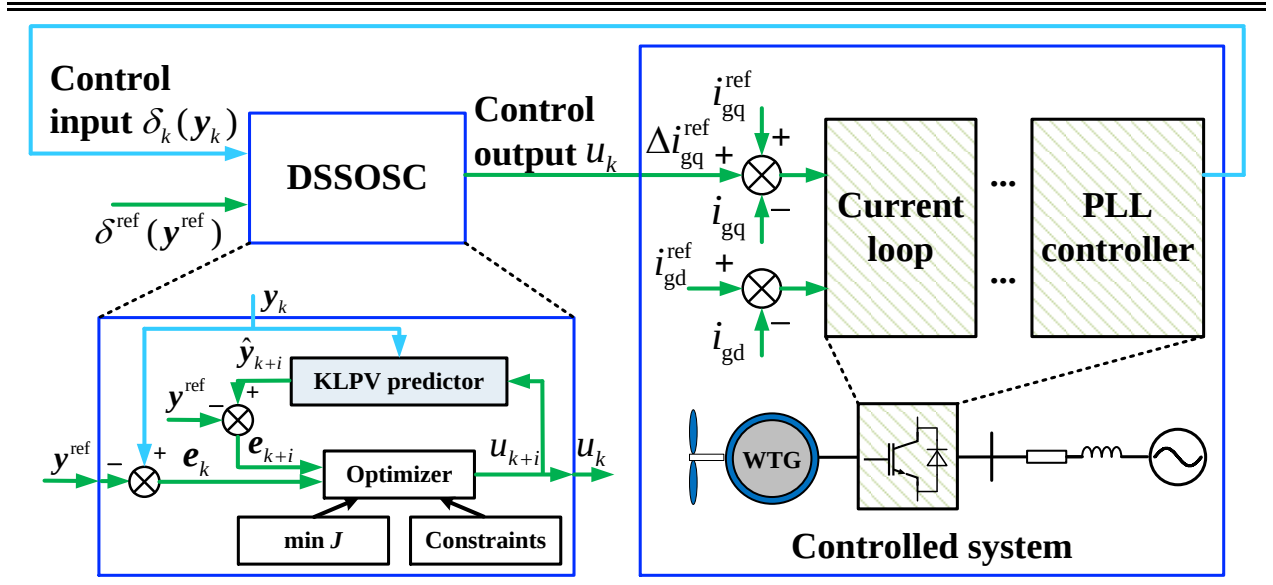

Fig. 5. Control scheme of DSSOSC attached to WTGs in REI power systems.

In the first REI test system, by combining datasets and the basis function, we perform Koopman mode decomposition via EDMD, and calculate the related information of modes with negative damping ratios. These modes manifest as divergent oscillations, corresponding to dominant modes of the REI power system in the poorly damped condition. As shown in Table I, the oscillation frequency of the Koopman eigenvalues is 10.74 Hz, within the frequency range of SSOs.

Fig. 4 displays the state-in-mode KPFs of the dominant mode (the SSO mode) computed using (11), where *PolyT*-Q denotes polynomials of order Q for the state variables [4]. From the KPFs results, the SSO mode has the top 7 lifted states as { $x_{iq}$, $\delta$, $\sin\delta$, $v_{dc}$, $\sin^2\delta$, $i_{gq}\cos\delta$, $\cos^2\delta$ }, and then $x_{iq}$ and $\delta$ are key state variables that influence the dominant mode of the REI power system. Furthermore, the participation degree of the controller in the SSO mode is quantified by summing up all state variables' KPFs internal to the same component [28]. The participation degree analysis of different components in the SSO mode of the REI power system is shown in Table II. At the controller level, the PLL controller and the current loop of the grid-side converter dominate the SSO mode. Hence, in the control scheme of DSSOSC, select the phase difference $\delta$ as the input signal and the increment of the current reference $i_{gq}^{ref}$ (directly affecting the state $x_{iq}$) as the output signal. The computed current reference increment $\Delta i_{gq}^{ref}$ through the KLPV MPC algorithm is fed back to the current control loop of the converters within WTGs. The DSSOSC to regulate WTGs with explicit control signals is formed in Fig. 5.

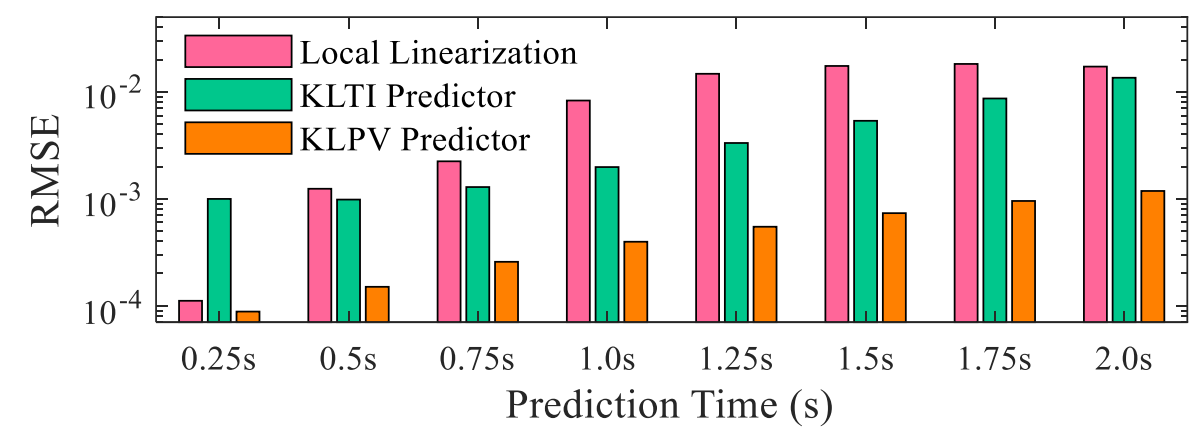

Fig. 6. Prediction RMSE with the increase in prediction time.

*2) Prediction Performance*

DSSOSC utilizes the KLPV predictor to identify controlled systems. The prediction performance of the controlled REI power system is compared with the local linearization and KLTI predictor in the typical SSO condition described in Section V-A. Fig. 6 reports the prediction inaccuracy quantified by the root mean squared error (RMSE). As the prediction time increases, the KLTI predictor tends to accumulate errors and brings a notable rise in prediction inaccuracies, with error values exceeding KLPV by up to 10 times. The prediction performance of local linearization deteriorates the fastest after the divergent SSO exceeds the small-signal range.

The KLPV predictor significantly outweighs the local linearization and KLTI predictor in prediction accuracy. The reason why the KLPV predictor is more accurate is that it takes into account the nonlinear characteristics of SSOs and the influence of inputs on the dynamics of the basis function during the Koopman linearization. Due to its ability to restrict the RMSE within 0.002 over a relatively long period of SSOs, the KLPV predictor is considered accurate enough to be applied in dynamics description and data-driven dynamic optimal control.

*3) Control Performance*

To evaluate the control performance of DSSOSC, the following three SSO suppression supplementary controllers are compared: the conventional supplementary damping controller (CSDC), KLTI MPC controller (KLTIC) and DSSOSC. The SSO phenomenon occurs after the grid structure changes at 0.5s, and the current reference $i_{gq}^{ref}$ of the converter current loop is adjusted by the supplementary controller to suppress oscillation. The principle of phase compensation is adopted for the design of CSDC [7], with the control objective of moving the eigenvalues of unstable modes towards the negative real axis to enhance mode damping. As a small-signal stability controller, the transfer function of CSDC based on phase compensation can be written as

$$H_{CSDC}(s) = K_p \frac{T_w s}{1+T_w s}\left(\frac{1+T_1 s}{1+T_2 s}\right)^2 \quad (28)$$

where $K_p$ is CSDC gain, $T_w$ is the wash-out constant, $T_1$ and $T_2$ are the lead/lag time constants, respectively. These parameters can be tuned by selecting appropriate eigenvalues on the root locus of the SSO mode as shown in Fig. 7.

Fig. 8 shows the control performance comparison of three SSO suppression controllers, with reference to an operating condition without controllers (denoted as W/OC). Parameters of CSDC are $K_p = 0.20$, $T_w = 0.10$, $T_1 = 0.1525$ and $T_2 = 0.0014$. KLTIC and DSSOSC adopt the same polynomial basis function up to third order, and the MPC parameters of them are the prediction horizon $N_p$ = 25 (50ms), and weight factors $\boldsymbol{Q}$ = 40, $\boldsymbol{R}$ = 0.01 with single input and single output. It can be observed that the system experiences a divergent SSO from 0.5s, and all three controllers are able to effectively suppress the SSO. However, DSSOSC outperforms CSDC and KLTIC in terms of the oscillation amplitude and settling time.

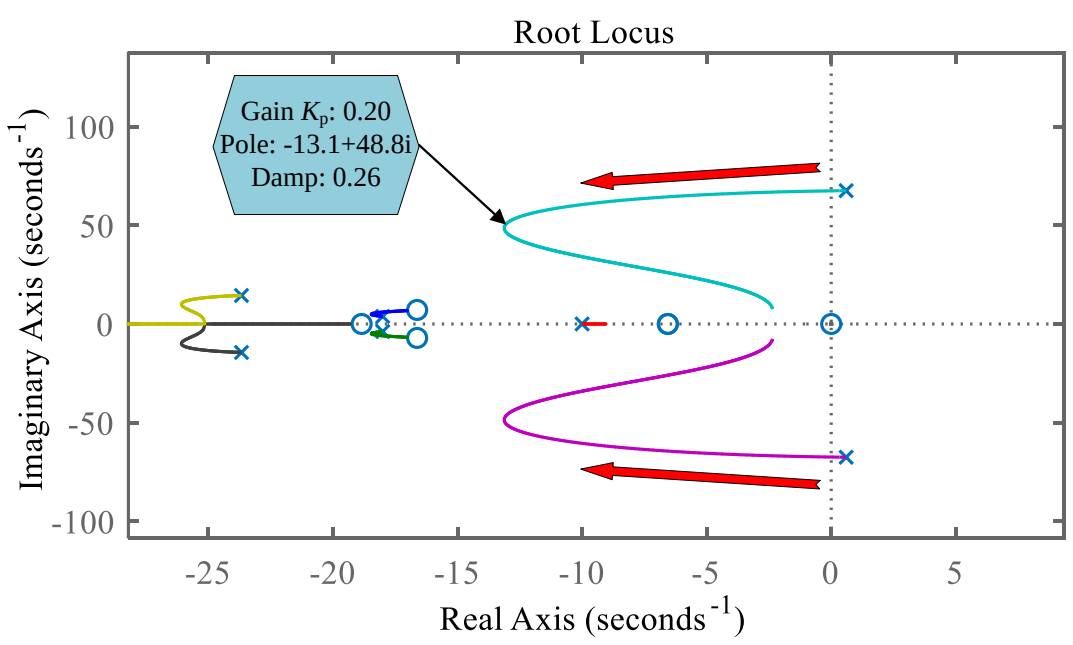

Fig. 7. Root locus of the dominant eigenvalues under SSO using CSDC.

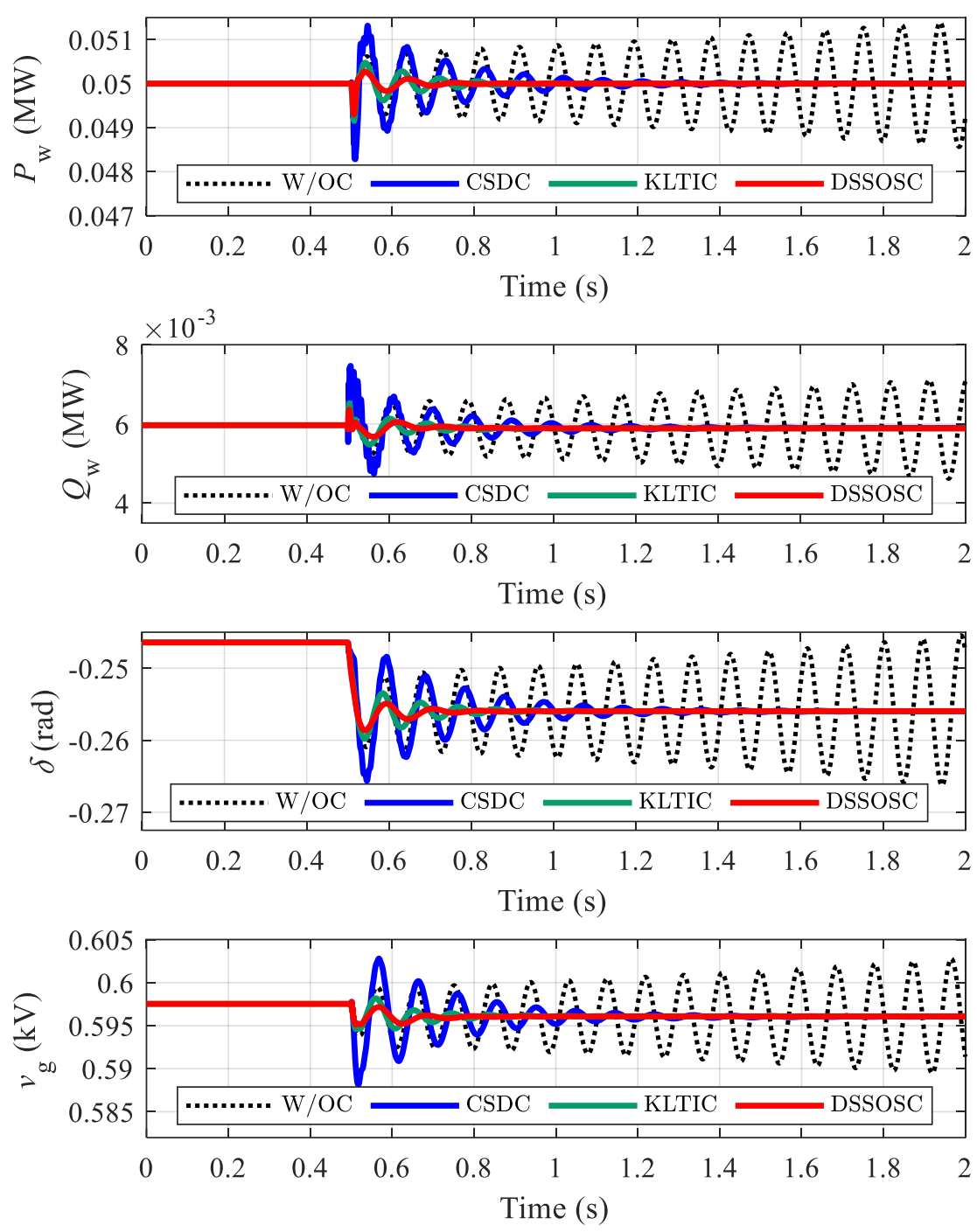

Fig. 8. Control performance comparison of three SSO suppression controllers.

TABLE IV
CONTROL PERFORMANCE INDEXES OF THE STUDIED CONTROLLERS

| Controller Type | Maximum Oscillation Amplitude of $P_w$ (kW) | Oscillation Settling Time (s) |
|---|---|---|
| CSDC | 1.7183 | 1.4564 |
| KLTIC | 0.8518 | 0.9602 |
| DSSOSC | 0.6993 | 0.5374 |

Table III quantifies the control performance of the studied controllers. As for the maximum oscillation amplitude of active power $P_w$ of a single WTG, the magnitude is regulated as 0.6993 kW using DSSOSC, while using KLTIC and CSDC, the magnitude is 1.22 times and 2.46 times that of DSSOSC, respectively. By analyzing the response curves of the active power $P_w$, the reactive power $Q_w$, the phase difference $\delta$ and the PCC voltage $v_g$, it is observed that the phase compensation in CSDC introduces a certain degree of overshoot, resulting in an increase in the oscillation amplitude. Additionally, in terms of the oscillation settling time, DSSOSC achieves oscillation suppression within 0.5374s, reducing the time by 44.03% and 63.10% compared to using KLTIC and using CSDC.

Therefore, in terms of the maximum oscillation amplitude and oscillation settling time, the control performance of DSSOSC is far superior to the other two controllers.

TABLE III
DOMINANT MODE OF OSCILLATION FROM KOOPMAN MODE DECOMPOSITION IN THE REI KUNDUR TWO-AREA POWER SYSTEM

| Mode Type | Eigenvalue | Frequency(Hz) | Damping Ratio(%) |
|---|---|---|---|
| SSO | 5.69±78.16i | 12.47 | -7.26 |

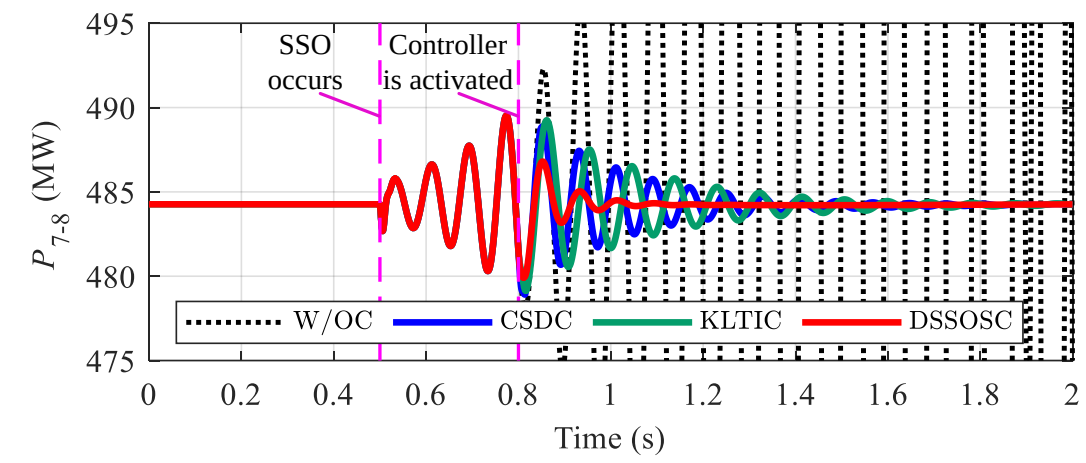

Fig. 9. Active power curves of tie line 7-8 with different controllers.

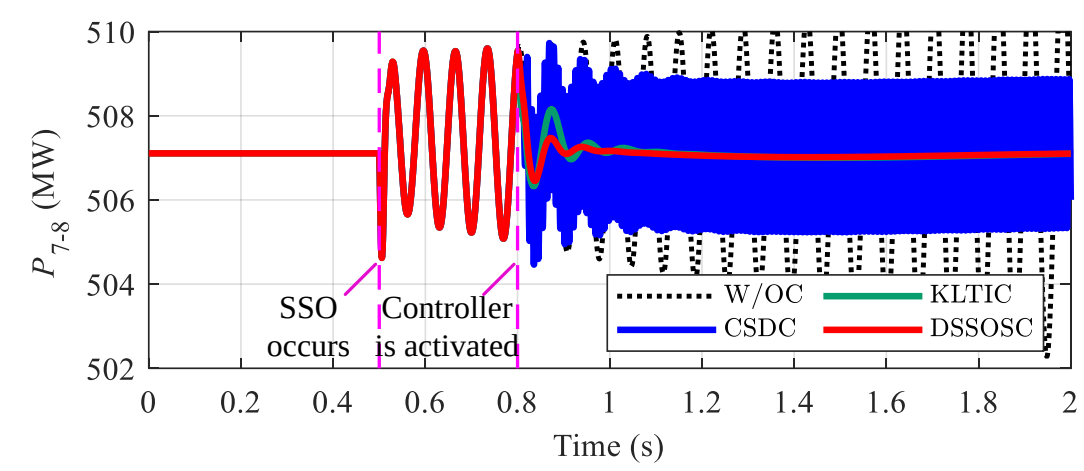

Fig. 10. Active power curves of tie line 7-8 when operating condition changes.

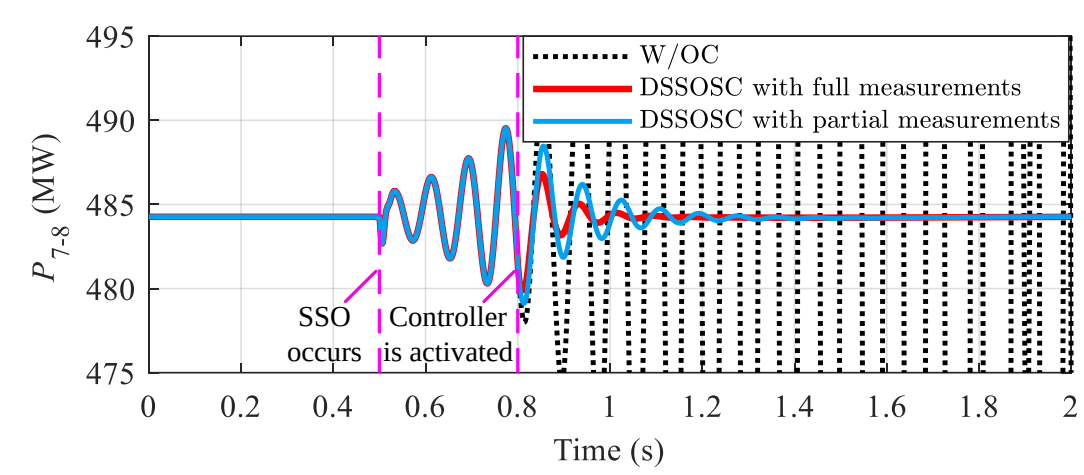

Fig. 11. Active power curves of tie line 7-8 considering the availability of full measurements or partial measurements.

## C. REI Kundur Two-Area Power System

To further evaluate the control performance of DSSOSC to external system changes, the simulation was also conducted on a modified Kundur two-area power system integrated with a wind farm. As a common benchmark used to analyze oscillations in interconnected power systems, the Kundur two-area power system faces the new challenge of SSOs when bulk RPGs are integrated via a long-distance transmission line.

The dominant mode information via Koopman mode decomposition from measurements are given in Table IV. The 12.47-Hz SSO corresponding to the dominant mode can be observed on the tie line from Bus-7 to Bus-8. The WTGs are identified as the key related generators of SSOs, and the DSSOSC is attached to the wind farm to enhance stability.

At 0.5s, an incorrect switch of the control parameter combinations in the converters within the wind farm occurs, resulting in the SSO, and after 0.3s, the SSO suppression controller is activated. As shown in Fig. 9, starting from the controller activation, DSSOSC achieves oscillation suppression within 0.3424s, only 0.28 times and 0.30 times the settling time of KLTIC and CSDC, respectively. As for the oscillation amplitude, DSSOSC achieves a reduction approximately half of the amplitude compared to other controllers, at around 0.86s. Thus, DSSOSC outperforms the compared controllers.

To study the adaptability of DSSOSC, we change the operating condition of the test system and increase the output of the wind farm. It is observed that the SSO frequency on the tie line from Bus-7 to Bus-8 changes accordingly. Fig. 10 shows that CSDC is no longer able to suppress SSOs, and even yields higher frequency oscillations, further endangering the system stability. KLTIC has a certain degree of adaptability; however, DSSOSC significantly outperforms it by reducing the oscillation settling time by 30.65%. DSSOSC maintains the best control performance. This is because the phase compensation parameters of CSDC are fixed, failing to match the phase compensation requirement of the different operating conditions. Meanwhile, DSSOSC, using the KLPV predictor with higher prediction accuracy, can identify the extended system online and adjust the control sequences adaptively, so as to suppress SSOs at different frequencies effectively.

To ensure the effectiveness and applicability of DSSOSC to real-world settings, the availability of measurements is considered. The previous analysis assumes that full state variables are measurable, but due to sensor limitations it is common to have limited access to measurements without using techniques like dynamic state estimators [26]. When partial measurements are available, only 12 variables, the phase difference $\delta_{\mathrm{w}}$, angular frequency $\omega_{\mathrm{w}}$, d-frame current $i_{\mathrm{gd}}$ and q-frame current $i_{\mathrm{gq}}$, as well as the SG related data $\delta_{\mathrm{G}j}$, $\omega_{\mathrm{G}j}$, ($j$=1,2,3,4), are selected to form the measurement dataset. Fig. 11 demonstrates that partial measurements lead to a slight increase in the maximum oscillation amplitude, but the control effectiveness is not affected. The reason lies in the fact that the KLPV predictor with the chosen observed variables can still effectively describe the essential dynamics of SSO mode, and DSSOSC possesses robustness against limited measurements.

Therefore, CSDC has a poor adaptability against different SSO frequencies, while DSSOSC exhibits advantages in suppressing SSOs at different frequencies more quickly. Due to its accurate state prediction ability, strong adaptability and advanced control algorithm, DSSOSC can effectively suppress SSOs even with partial measurements, showcasing better and more reliable control performance.

### *D. Large-scale Power System Using Partial Measurements*

To test the proposed DSSOSC in a large-scale power system, the modified REI IEEE 39-bus system was studied. In this system as illustrated in Fig. 12, SGs connected from Bus-30 to Bus-36 are substituted with grid-following converter based RPGs like WTGs and PVs. The parameters of these RPGs are illustrated in Appendix C. All SGs including G8, G9 and G10 are equipped with power system stabilizers (PSSs) in the excitation system and G10 at Bus-39 is taken as a reference. The WTGs in the wind farm G1 at Bus-30 transmit power to Bus-2 via a long-distance double-circuit line, and due to maintenance resulting from line icing, one of the double-circuit lines is cut off. These factors cause a weak grid disturbance and lead to SSOs. DSSOSC has been verified to maintain effective control performance in Section V-C with partial measurements. The same partial measurements for generators are collected, including PLL angle $\delta_{\mathrm{R}l}$, angular frequency $\omega_{\mathrm{R}l}$, d-frame current $i_{\mathrm{gd}l}$ and q-frame current $i_{\mathrm{gq}l}$ of RPGs ($l$=1,2,…,7), as well as rotor angle $\delta_{\mathrm{G}j}$ and angular frequency $\omega_{\mathrm{G}j}$ of SGs ($j$=8,9,10). The basis functions are created by *PolyT*-2 [4]. The dimension of basis functions for each RPG is 21 and for each SG is 9.

Since the dominant mode depicted in Table V is negatively damped under the weak grid disturbance, the transient response without controllers shows the characteristic of divergent SSO. The Koopman mode decomposition can describe the accurate dominant oscillation mode in the large-scale power system. The 18.77-Hz SSO is observed in the active power of the tie-line from Bus-39 to Bus-9 and relative angles (rotor angles of SGs or PLL angles of RPGs) as depicted in Fig. 13(a).

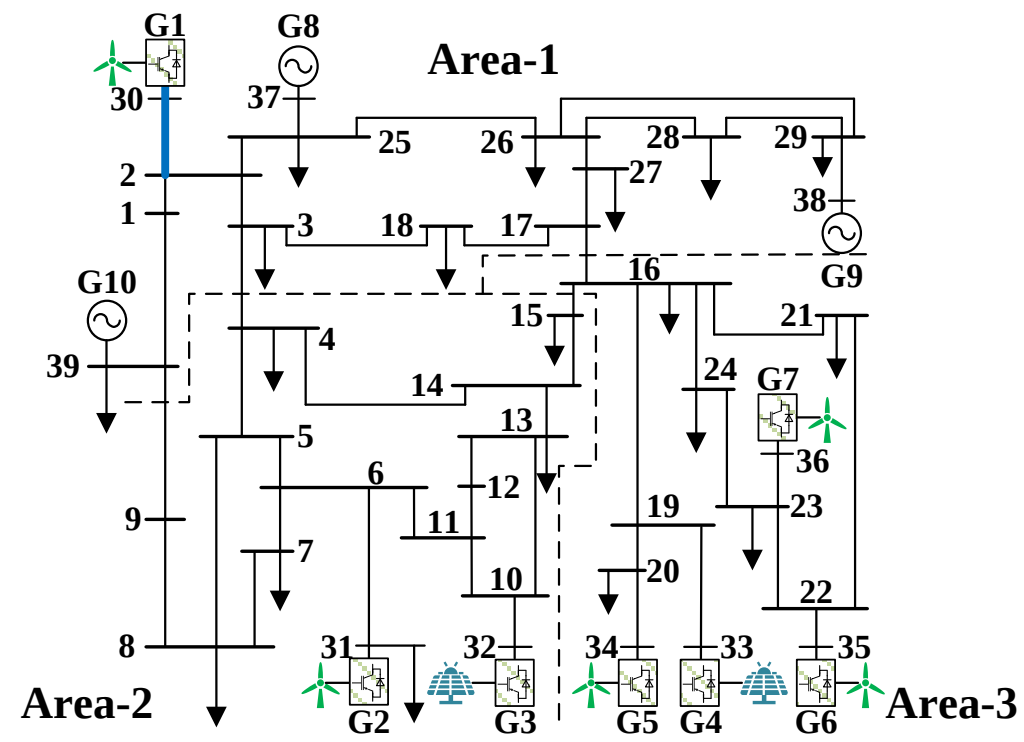

Fig. 12. Diagram of the modified REI IEEE 39-bus system.

TABLE V
DOMINANT MODE OF OSCILLATION FROM KOOPMAN MODE DECOMPOSITION IN THE REI IEEE 39-BUS SYSTEM

| Mode Type | Eigenvalue | Frequency(Hz) | Damping Ratio(%) |
|---|---|---|---|
| SSO | 1.33±117.95i | 18.77 | -1.13 |

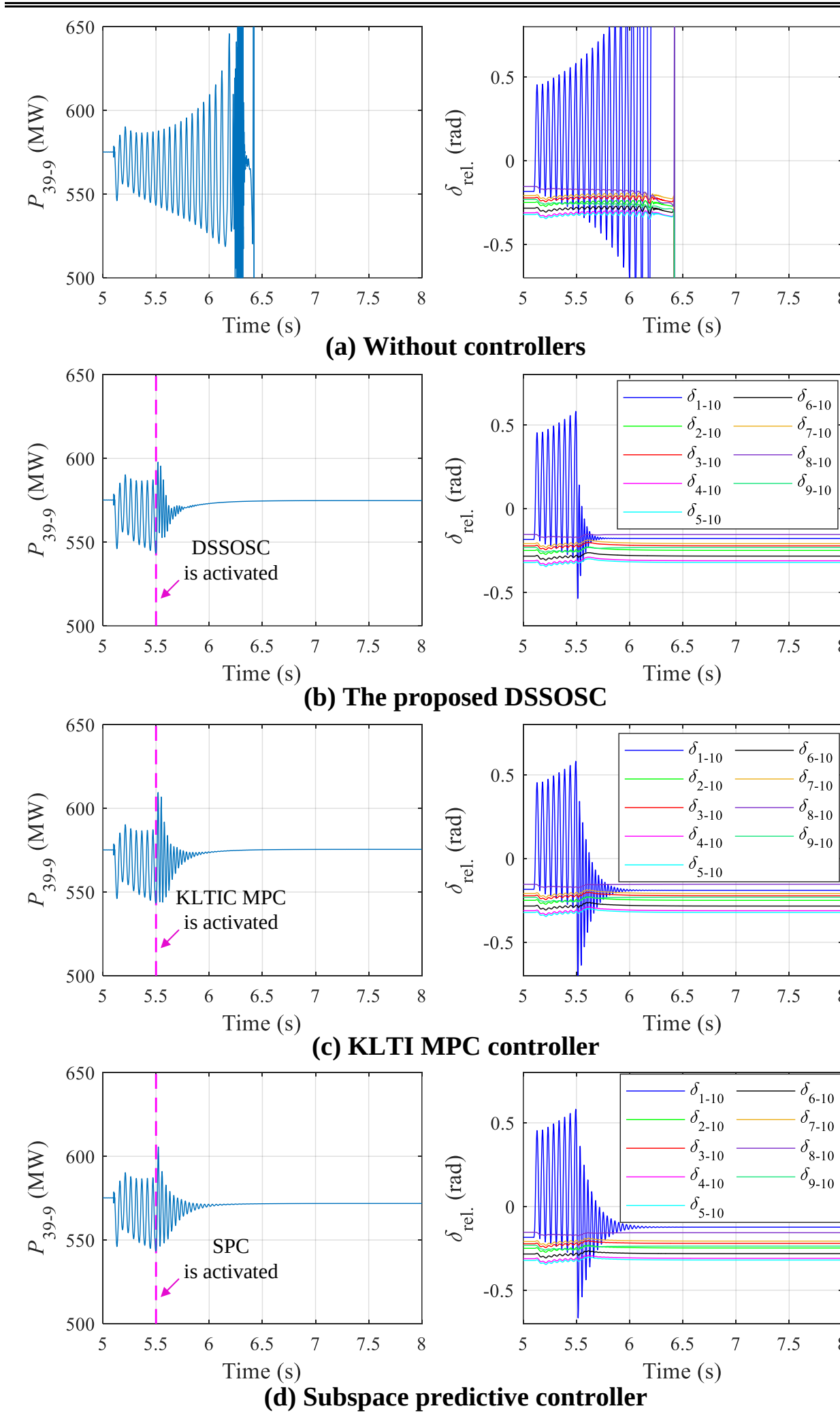

Fig. 13. Active power of tie line 39-9 and relative (rotor or PLL) angle curves without controllers and with different SSO suppression controllers.

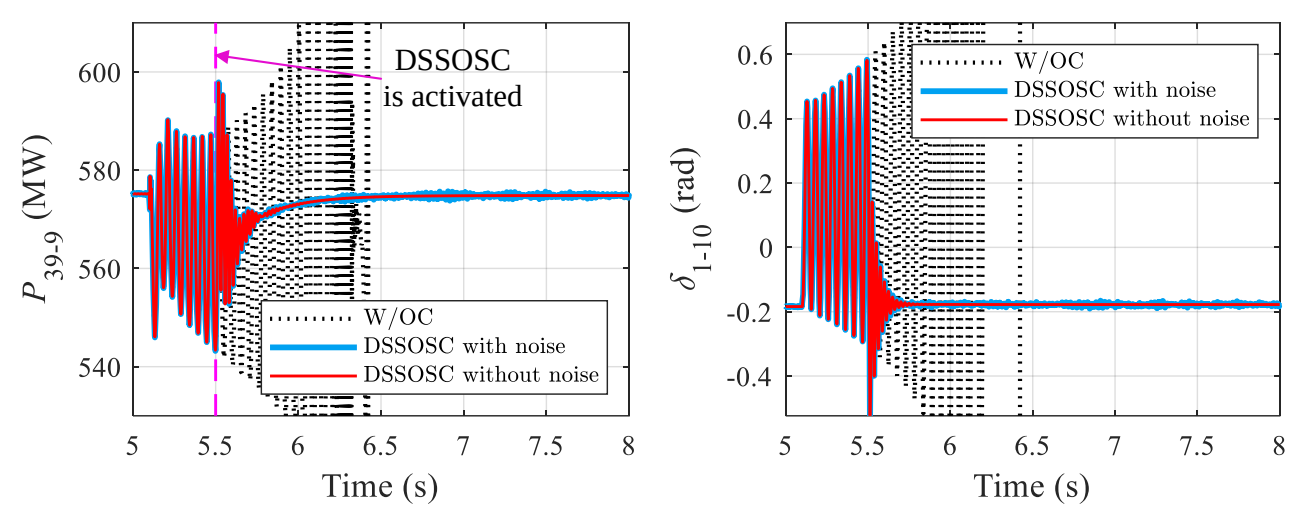

Fig. 14. Active power of tie line 39-9 and relative angle $\delta_{1\text{-}10}$ with 40 dB noise.

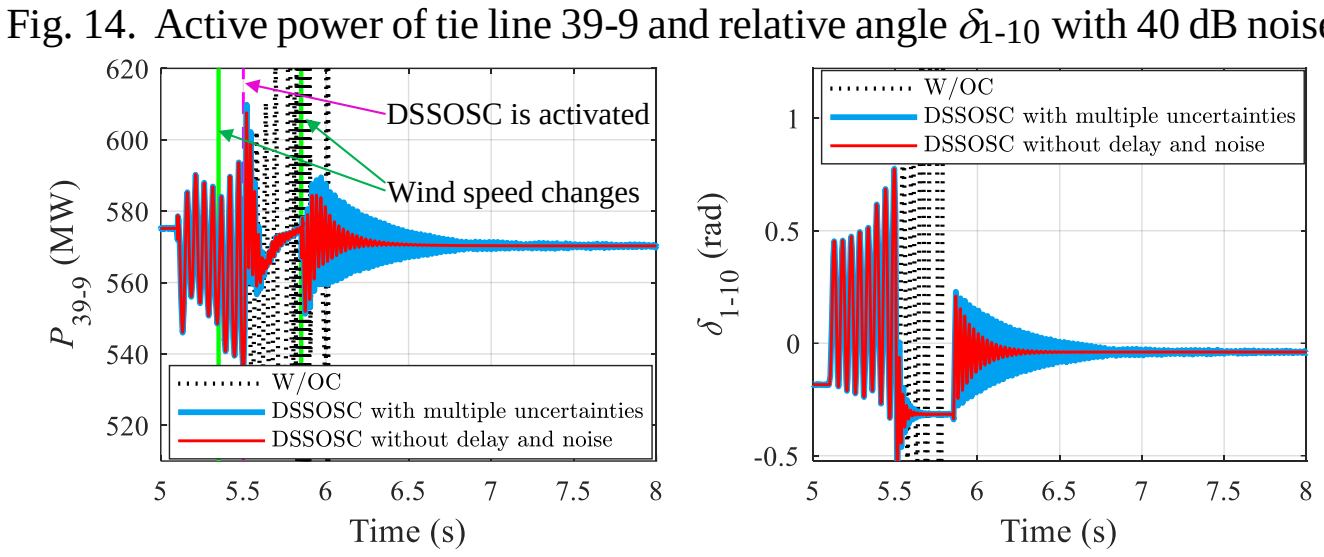

Fig. 15. Active power of tie line 39-9 and relative angle $\delta_{1\text{-}10}$ with multiple uncertainties including varying wind speed, noise, and communication delay.

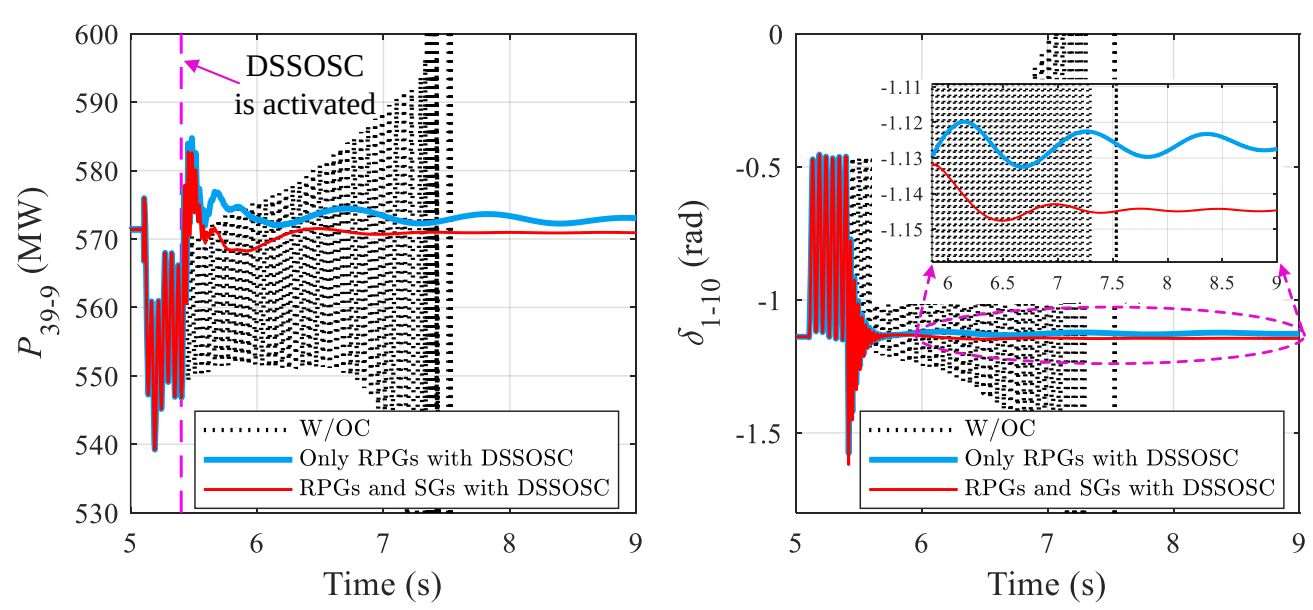

Fig. 16. Active power of tie line 39-9 and relative angle $\delta_{1\text{-}10}$ under multiband oscillations in the operating condition where SGs are not equipped with PSSs.

In the context of engineering practice, the control superiority of DSSOSC is verified under the following conditions:

a) *Different SSO suppression controllers*: Compared to the unstable oscillation occurring at 5.1s, Fig. 13 conveys that DSSOSC can suppress the SSO effectively when the controller is activated at 5.5s. A novel oscillation suppression controller using subspace predictive control [29], [30] has also been introduced for comparison. In contrast to KLTI MPC controller and subspace predictive controller (SPC), DSSOSC can decrease the oscillation settling time by 36.34% and 49.12% respectively. Due to its best ability to characterize nonlinearity, DSSOSC possesses the superior control performance, followed by KLTI MPC. SPC is not suitable for nonlinear systems as its controller design is wholly founded on linear predictive models, resulting in inferior control performance.

b) *Measurement noise*: In the context of engineering practice, the robustness of DSSOSC is verified towards measurement noise, uncertainties of RPGs and communication delay. To simulate the possible interference found in real-world measurement data, we inject 40dB random Gaussian noise signals into the sampled data. Fig. 14 indicates that although measurement noise may lead to slight fluctuations in the tie-line power $P_{39\text{-}9}$ and relative angle $\delta_{1\text{-}10}$, DSSOSC can maintain reliable control performance and is robust to certain level of measurement noise for power systems.

c) *Multiple uncertainties.* To further validate the robustness, uncertainties in wind speed and communication delay [31] are considered in addition to the 40dB measurement noise. The wind speed of 9m/s increases at 5.35s and decreases at 5.85s with a fluctuation range of ±10%, and communication delay is 50ms. Numerous tests show that DSSOSC has inherent robustness to suppress the SSO against the communication delay of at least 15ms, and mature techniques such as delay compensation [32], [33], [34] for MPC can be employed in DSSOSC to further mitigate the effects of delays. Fig. 15 demonstrates that multiple uncertainties lead to a slight degradation in control performance; however, the SSO is suppressed by DSSOSC successfully, and the settling time is guaranteed to be within 2s. Thus, DSSOSC possesses strong robustness against these practical uncertainties.

d) *Multiband oscillations*: In the former operating condition, due to the installation of PSSs in all SGs, only the SSO mode shows up and can be suppressed by DSSOSC. When the operating condition changes and the SGs are not equipped with PSSs, the system disturbance triggers not only the SSO mode but also the low-frequency electromechanical mode. In this condition, the dominant generator for the SSO mode is the wind farm G1, and the dominant generators for the low-frequency oscillation mode are SGs G8, G9, and G10. As shown in Fig. 16, installing the controller DSSOSC in RPGs can suppress the SSO mode dominated by RPGs. In addition, more DSSOSCs are installed in the excitation system of SGs, and control input and output signals are the same as the control signals of PSSs, which is capable of suppressing the low-frequency oscillation dominated by SGs. This indicates that the proposed DSSOSC can effectively suppress the low-frequency oscillation and SSO, two types of multiband oscillations, and can be applied to different types of generators. Therefore, the DSSOSC is a data-driven oscillation suppression controller that is not confined to fixed oscillation frequency band and controlled objects, and boasts significant adaptability and scalability.

## VI. Conclusion

In this paper, a data-driven controller DSSOSC is developed for REI power systems to suppress SSOs with the control of RPGs. Based on KO, by analyzing the KPFs, constructing the KLPV predictor, and utilizing the KLPV MPC algorithm, we sequentially implement the signal selection, state prediction, and linear control utilization. The advantage of KO mainly lies in its ability to consider the impact of nonlinearity in the SSO mode and to provide a global linear representation. The global linear representation facilitates the computationally efficient linear MPC of nonlinear dynamical systems. In addition, the proposed KLPV MPC with enhanced prediction accuracy (reducing errors by at least 80%) can ensure better and more reliable control performance for DSSOSC.

Case studies demonstrate the effectiveness, adaptability and robustness of the proposed data-driven controller DSSOSC in suppressing SSOs under different practical conditions even with varying oscillation frequencies, partial measurements, noise, uncertainties of RPGs and communication delay. Due to its accurate state prediction ability and advanced control algorithm, DSSOSC can avoid the reliance on detailed system models and suppress SSOs with at least 30.65% reduction in the settling time. Therefore, DSSOSC integrates the data-driven method into the stability enhancement control, offering broad application prospects in practical large-scale power systems with high penetrations of renewable power generation.

## APPENDIX

### A. Proof of the Eigenfunction Form in the Observable Space

Given the numerical approximations of eigenfunctions $\boldsymbol{\varphi}(\boldsymbol{x}_k)$ in (9) from the EDMD algorithm, and the evolution property $\boldsymbol{\gamma}(\boldsymbol{x}_k)=\boldsymbol{K}_{\mathrm{d}}\boldsymbol{\gamma}(\boldsymbol{x}_{k-1})$ of $\boldsymbol{K}_{\mathrm{d}}$ which acts in the observable space, eigenfunctions have the following recursive relation

$$\boldsymbol{\varphi}(\boldsymbol{x}_k)=\boldsymbol{\Xi}\boldsymbol{\gamma}(\boldsymbol{x}_k)=\boldsymbol{\Xi}\boldsymbol{K}_{\mathrm{d}}\boldsymbol{\gamma}(\boldsymbol{x}_{k-1})=\boldsymbol{\Xi}\boldsymbol{K}_{\mathrm{d}}\boldsymbol{\Xi}^{-1}\boldsymbol{\varphi}(\boldsymbol{x}_{k-1}) \quad (29)$$

Because $\boldsymbol{\Xi}$ and $\mathbf{U}$, matrices containing left and right eigenvectors, satisfy the relations $\boldsymbol{\Xi}\boldsymbol{K}_{\mathrm{d}}=\boldsymbol{\Lambda}\boldsymbol{\Xi}$, $\boldsymbol{K}_{\mathrm{d}}\mathbf{U}=\mathbf{U}\boldsymbol{\Lambda}$ and $\boldsymbol{\Xi}\mathbf{U}=\mathbf{I}$, the eigendecomposition of $\boldsymbol{K}_{\mathrm{d}}$ is of the form

$$\boldsymbol{K}_{\mathrm{d}}=\mathbf{U}\boldsymbol{\Lambda}\boldsymbol{\Xi}=\boldsymbol{\Xi}^{-1}\boldsymbol{\Lambda}\boldsymbol{\Xi} \quad (30)$$

where $\boldsymbol{\Lambda}$ is the diagonal matrix composed of eigenvalues $\mu_i$.

Then, (29) can be expressed as

$$\boldsymbol{\varphi}(\boldsymbol{x}_k)=\boldsymbol{\Lambda}\boldsymbol{\varphi}(\boldsymbol{x}_{k-1})=\boldsymbol{\Lambda}^k\boldsymbol{\varphi}(\boldsymbol{x}_0) \quad (31)$$

Therefore, the $i$-th eigenfunction obeys the evolution form in the observable space

$$\varphi_i(\boldsymbol{x}_k)=\mu_i^k\varphi_i(\boldsymbol{x}_0)=\mu_i^k\boldsymbol{\xi}_i\boldsymbol{\gamma}(\boldsymbol{x}_0)=\mu_i^k\sum_{j=1}^{m}\xi_{ij}\boldsymbol{\gamma}(\boldsymbol{x}_0) \quad (32)$$

### B. Dynamics of the REI Power System with Weak Grid SSOs

The dynamics of the REI power system can be written as

a) DC link model

$$\dot{v}_{\mathrm{dc}}=\frac{1}{C_{\mathrm{dc}}}\left(i_{\mathrm{sc}}-\frac{3}{2v_{\mathrm{dc}}}\left(v_{\mathrm{od}}i_{\mathrm{rd}}+v_{\mathrm{oq}}i_{\mathrm{rq}}\right)\right) \quad (33)$$

b) Voltage loop model

$$\dot{x}_{\mathrm{v}}=v_{\mathrm{dc}}^{\mathrm{ref}}-v_{\mathrm{dc}} \quad (34)$$

$$i_{\mathrm{gd}}^{\mathrm{ref}}=k_{\mathrm{P1}}\left(v_{\mathrm{dc}}^{\mathrm{ref}}-v_{\mathrm{dc}}\right)+k_{\mathrm{I1}}x_{\mathrm{v}}\ ,\ i_{\mathrm{gq}}^{\mathrm{ref}}=0 \quad (35)$$

c) Current loop model

$$\begin{bmatrix}\dot{x}_{\mathrm{id}}\\ \dot{x}_{\mathrm{iq}}\end{bmatrix}=\begin{bmatrix}i_{\mathrm{gd}}^{\mathrm{ref}}\\ i_{\mathrm{gq}}^{\mathrm{ref}}\end{bmatrix}-\begin{bmatrix}i_{\mathrm{gd}}\\ i_{\mathrm{gq}}\end{bmatrix} \quad (36)$$

$$\begin{bmatrix}\dot{v}_{\mathrm{od}}\\ \dot{v}_{\mathrm{oq}}\end{bmatrix}=\frac{1}{T_{\mathrm{d}}}\left(\begin{bmatrix}v_{\mathrm{od}}^{\mathrm{ref}}\\ v_{\mathrm{oq}}^{\mathrm{ref}}\end{bmatrix}-\begin{bmatrix}v_{\mathrm{od}}\\ v_{\mathrm{oq}}\end{bmatrix}\right) \quad (37)$$

$$\begin{bmatrix}v_{\mathrm{od}}^{\mathrm{ref}}\\ v_{\mathrm{oq}}^{\mathrm{ref}}\end{bmatrix}=k_{\mathrm{P2}}\left(\begin{bmatrix}i_{\mathrm{gd}}^{\mathrm{ref}}\\ i_{\mathrm{gq}}^{\mathrm{ref}}\end{bmatrix}-\begin{bmatrix}i_{\mathrm{gd}}\\ i_{\mathrm{gq}}\end{bmatrix}\right)+k_{\mathrm{I2}}\begin{bmatrix}x_{\mathrm{id}}\\ x_{\mathrm{iq}}\end{bmatrix} \quad (38)$$

d) PLL controller model

$$\dot{x}_{\omega}=v_{\mathrm{gq}} \quad (39)$$

$$\dot{\delta}=\omega-\omega_0=k_{\mathrm{P_pll}}v_{\mathrm{gq}}+k_{\mathrm{I_pll}}x_{\omega} \quad (40)$$

e) LCL filter and the network model

$$\begin{bmatrix}\dot{i}_{\mathrm{rd}}\\ \dot{i}_{\mathrm{rq}}\end{bmatrix}=\frac{1}{L_{\mathrm{r}}}\left(\begin{bmatrix}v_{\mathrm{od}}\\ v_{\mathrm{oq}}\end{bmatrix}-\begin{bmatrix}v_{\mathrm{rd}}\\ v_{\mathrm{rq}}\end{bmatrix}-\begin{bmatrix}R_{\mathrm{r}} & -\omega L_{\mathrm{r}}\\ \omega L_{\mathrm{r}} & R_{\mathrm{r}}\end{bmatrix}\begin{bmatrix}i_{\mathrm{rd}}\\ i_{\mathrm{rq}}\end{bmatrix}\right) \quad (41)$$

$$\begin{bmatrix}\dot{v}_{\mathrm{cd}}\\ \dot{v}_{\mathrm{cq}}\end{bmatrix}=\frac{1}{C_{\mathrm{r}}}\left(\begin{bmatrix}i_{\mathrm{rd}}\\ i_{\mathrm{rq}}\end{bmatrix}-\begin{bmatrix}i_{\mathrm{gd}}\\ i_{\mathrm{gq}}\end{bmatrix}-\begin{bmatrix}0 & -\omega C_{\mathrm{r}}\\ \omega C_{\mathrm{r}} & 0\end{bmatrix}\begin{bmatrix}v_{\mathrm{cd}}\\ v_{\mathrm{cq}}\end{bmatrix}\right) \quad (42)$$

$$\begin{bmatrix}\dot{i}_{\mathrm{gd}}\\ \dot{i}_{\mathrm{gq}}\end{bmatrix}=\frac{1}{L_{\Sigma}}\left(\begin{bmatrix}v_{\mathrm{rd}}\\ v_{\mathrm{rq}}\end{bmatrix}-\begin{bmatrix}v_{\mathrm{sd}}\\ v_{\mathrm{sq}}\end{bmatrix}-\begin{bmatrix}R_{\Sigma} & -\omega L_{\Sigma}\\ \omega L_{\Sigma} & R_{\Sigma}\end{bmatrix}\begin{bmatrix}i_{\mathrm{gd}}\\ i_{\mathrm{gq}}\end{bmatrix}\right) \quad (43)$$

$$\begin{bmatrix}v_{\mathrm{sd}}\\ v_{\mathrm{sq}}\end{bmatrix}=\begin{bmatrix}v_{\mathrm{sm}}\cos\delta\\ -v_{\mathrm{sm}}\sin\delta\end{bmatrix} \quad (44)$$

$$\begin{bmatrix}v_{\mathrm{rd}}\\ v_{\mathrm{rq}}\end{bmatrix}=\begin{bmatrix}v_{\mathrm{cd}}\\ v_{\mathrm{cq}}\end{bmatrix}+R_{\mathrm{c}}\left(\begin{bmatrix}i_{\mathrm{rd}}\\ i_{\mathrm{rq}}\end{bmatrix}-\begin{bmatrix}i_{\mathrm{gd}}\\ i_{\mathrm{gq}}\end{bmatrix}\right) \quad (45)$$

$$\begin{bmatrix}v_{\mathrm{gd}}\\ v_{\mathrm{gq}}\end{bmatrix}=\frac{L_{\mathrm{s}}L_{\mathrm{g}}}{L_{\Sigma}}\left(\frac{1}{L_{\mathrm{s}}}\begin{bmatrix}v_{\mathrm{sd}}\\ v_{\mathrm{sq}}\end{bmatrix}+\frac{1}{L_{\mathrm{g}}}\begin{bmatrix}v_{\mathrm{rd}}\\ v_{\mathrm{rq}}\end{bmatrix}-\left(\frac{r_{\mathrm{g}}}{L_{\mathrm{g}}}-\frac{r_{\mathrm{s}}}{L_{\mathrm{s}}}\right)\begin{bmatrix}i_{\mathrm{gd}}\\ i_{\mathrm{gq}}\end{bmatrix}\right) \quad (46)$$

where the state vector $\boldsymbol{x}$ : $[v_{\mathrm{dc}}\ x_{\mathrm{v}}\ x_{\mathrm{id}}\ x_{\mathrm{iq}}\ v_{\mathrm{od}}\ v_{\mathrm{oq}}\ x_{\omega}\ \delta\ i_{\mathrm{rd}}\ i_{\mathrm{rq}}\ v_{\mathrm{cd}}\ v_{\mathrm{cq}}\ i_{\mathrm{gd}}\ i_{\mathrm{gq}}]^{\mathrm{T}}$ has been explained in Section IV. Besides, $v_{\mathrm{rd}}$ and $v_{\mathrm{rq}}$ are the voltage variables of the capacitance $C_{\mathrm{r}}$ and its damping resistance $R_{\mathrm{c}}$. $\omega$ and $\omega_0$ are the output frequency of the PLL and frequency of the grid voltage $v_s$, respectively. $v_{\mathrm{dc}}^{\mathrm{ref}}$, $i_{\mathrm{gd}}^{\mathrm{ref}}$, $i_{\mathrm{gq}}^{\mathrm{ref}}$, $v_{\mathrm{od}}^{\mathrm{ref}}$ and $v_{\mathrm{oq}}^{\mathrm{ref}}$ are the references for the state variables in the voltage loop and current loop. $k_{\mathrm{P1}}$, $k_{\mathrm{I1}}$, $k_{\mathrm{P2}}$, $k_{\mathrm{I2}}$, $k_{\mathrm{P_pll}}$ and $k_{\mathrm{I_pll}}$ are the proportional and integral gains in the voltage loop, current loop and PLL controller, respectively. $T_{\mathrm{d}}$ is the time delay of the slow-scale dynamics. Furthermore, $C_{\mathrm{dc}}$, $L_{\mathrm{s}}$ and $R_{\mathrm{s}}$ are the dc-link capacitor, grid inductance and resistance, respectively. $R_{\mathrm{r}}$ and $R_{\mathrm{g}}$ are the stray resistances of the inductances $L_{\mathrm{r}}$ and $L_{\mathrm{g}}$ in the LCL filter. $R_{\Sigma}$ equals $R_{\mathrm{g}}$ plus $R_{\mathrm{s}}$, and $L_{\Sigma}$ equals $L_{\mathrm{g}}$ plus $L_{\mathrm{s}}$.

The weak AC grid is represented by a stiff voltage $v_{\mathrm{s}}$ with a large grid impedance connected in series, and the impedance is modeled as a series connection of $L_{\mathrm{s}}$ and $R_{\mathrm{s}}$. Algebraic equations (44) to (46) characterize the electrical interaction between the RPG and the power grid, encompassing pivotal details such as the power grid voltage $v_{\mathrm{s}}$, the PCC voltage $v_{\mathrm{g}}$, and the PLL angle $\delta$. Equations (33) to (46) represent the dynamic model of the REI power system with weak grid SSOs.

### C. Parameters of WTGs and PVs

The control of grid-following converter based WTGs mainly consists of the machine-side controller (MSC) and the grid-side controller (GSC). The parameters of WTGs in wind farms G2/5/6/7 are the same, as shown in Appendix Table I:

APPENDIX TABLE I
PARAMETERS OF THE WTGS IN WIND FARMS G2/5/6/7

| Symbols | Descriptions | Values |
|---|---|---|
| $k_{\mathrm{P1_MSC}}$ | proportional gain in the active power loop | 1 |
| $k_{\mathrm{I1_MSC}}$ | integral gain in the active power loop | 20 |
| $k_{\mathrm{P2_MSC}}$ | proportional gain in the MSC d-frame current loop | 1.5 |
| $k_{\mathrm{I2_MSC}}$ | integral gain in the MSC d-frame current loop | 30 |
| $k_{\mathrm{P3_MSC}}$ | proportional gain in the MSC q-frame current loop | 1.5 |
| $k_{\mathrm{I3_MSC}}$ | integral gain in the MSC q-frame current loop | 30 |
| $k_{\mathrm{P1_GSC}}$ | proportional gain in the voltage loop | 0.8 |
| $k_{\mathrm{I1_GSC}}$ | integral gain in the voltage loop | 10 |
| $k_{\mathrm{P2_GSC}}$ | proportional gain in the GSC d-frame current loop | 1 |
| $k_{\mathrm{I2_GSC}}$ | integral gain in the GSC d-frame current loop | 33 |
| $k_{\mathrm{P3_GSC}}$ | proportional gain in the GSC q-frame current loop | 1 |
| $k_{\mathrm{I3_GSC}}$ | integral gain in the GSC q-frame current loop | 33 |
| $k_{\mathrm{P_pll}}$ | proportional gain in the PLL controller | 50 |
| $k_{\mathrm{I_pll}}$ | integral gain in the PLL controller | 1000 |
| $v_{\mathrm{wind}}$ | wind speed | 9m/s |

The control parameters of WTGs in wind farm G1 are the same as other wind farms, except the integral gain in the PLL: $k_{I_pll}$=6500.

The control of grid-following converter based PVs is mainly determined by the GSC. The parameters of PVs in PV plants G3/4 are the same, as shown in Appendix Table II:

APPENDIX TABLE II
PARAMETERS OF THE PVS IN PV PLANTS G3/4

| Symbols | Descriptions | Values |
|---|---|---|
| $k_{P1_GSC}$ | proportional gain in the voltage loop | 0.75 |
| $k_{I1_GSC}$ | integral gain in the voltage loop | 10 |
| $k_{P2_GSC}$ | proportional gain in the GSC d-frame current loop | 0.5 |
| $k_{I2_GSC}$ | integral gain in the GSC d-frame current loop | 60 |
| $k_{P3_GSC}$ | proportional gain in the GSC q-frame current loop | 0.5 |
| $k_{I3_GSC}$ | integral gain in the GSC q-frame current loop | 60 |
| $k_{P_pll}$ | proportional gain in the PLL controller | 50 |
| $k_{I_pll}$ | integral gain in the PLL controller | 1000 |
| $S_{PV}$ | strength of illumination | 1000W/m$^2$ |
| $T_{PV}$ | temperature | 25℃ |

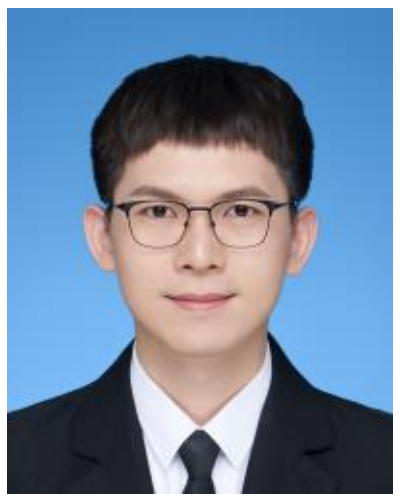

**Zihan Wang** (S’19-M’24) received the B.S. degree from North China Electric Power University, Beijing, China, in 2019, where he is currently working toward the Ph.D. degree in electrical engineering. His research interests include stability and control of power electronics dominated power systems and renewable energy conversion systems.

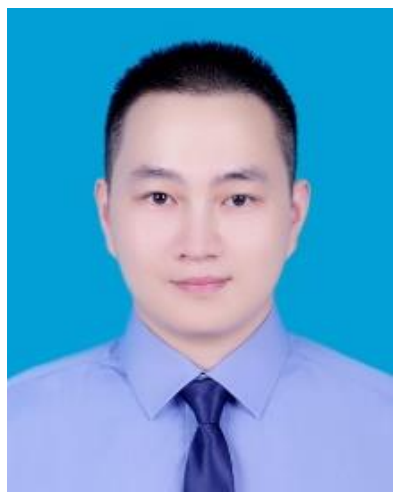

**Ziyang Huang** (S’23-M’24) received the B.S. degree in Electrical Engineering from North China Electric Power University, Beijing, China, in 2017, where he is currently pursuing the Ph.D. degree. His research interests include renewable energy planning involving voltage stability, and dynamic frequency regulation using wind turbine generator and energy storage.

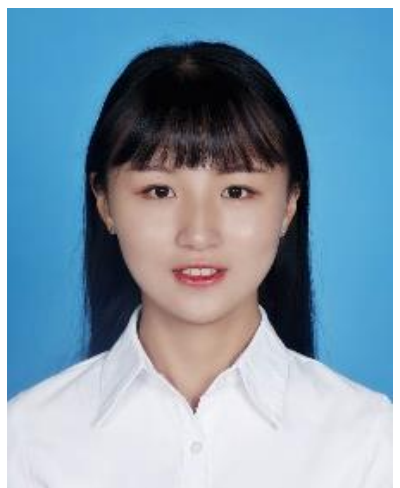

**Xiaonan Zhang** (S’21) received the B.S. degree in Electrical Engineering from North China Electric Power University, Beijing, China, in 2021. She is currently working toward the Ph.D. degree in Electrical Engineering at North China Electric Power University, Beijing, China. Her main research interest is the stability analysis and control of renewable energy power systems.

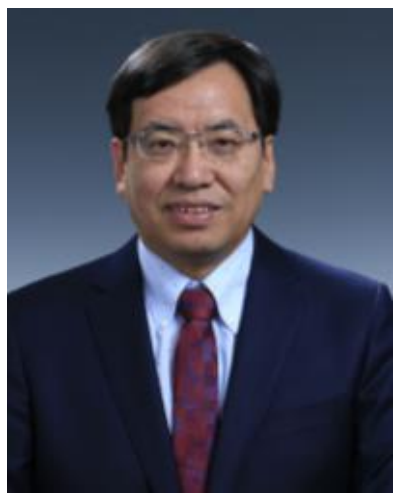

**Gengyin Li** (M’03) received his B.S., M.S., and Ph.D. degrees from North China Electric Power University, Beijing, China, in 1984, 1987, and 1996 respectively, all in electrical engineering. Since 1987, he has been with the School of Electrical and Electronic Engineering, North China Electric Power University, where he is currently a Professor. His research interests include HVDC transmission, power quality analysis and control, and emerging transmission and distribution technologies.

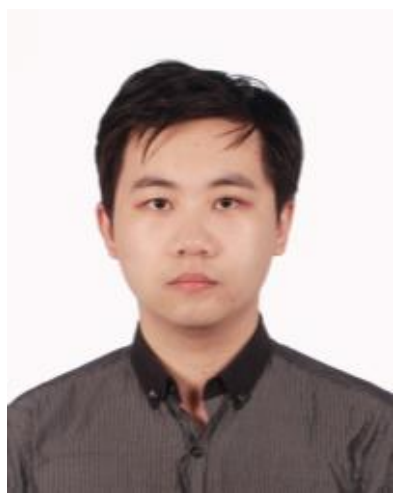

**Le Zheng** (M’12) received the B.S. and Ph.D. degrees in electrical engineering from Tsinghua University, Beijing, China, in 2011 and 2017, respectively. He had been a Postdoctoral Research Fellow in Stanford University from 2017 to 2019. He is currently an Associate Professor of the North China Electric Power University, Beijing, China. His current research interests include stability and control of power electronics dominated power systems and renewable energy conversion systems, and machine learning applications in power systems.